\documentclass{article}

\usepackage{arxiv}

\usepackage{amsmath}
\usepackage[utf8]{inputenc} % allow utf-8 input
\usepackage[T1]{fontenc}    % use 8-bit T1 fonts
\usepackage{hyperref}       % hyperlinks
\usepackage{url}            % simple URL typesetting
\usepackage{booktabs}       % professional-quality tables
\usepackage{amsfonts}       % blackboard math symbols
\usepackage{nicefrac}       % compact symbols for 1/2, etc.
\usepackage{microtype}      % microtypography
\usepackage{cleveref}       % smart cross-referencing
\usepackage{lipsum}         % Can be removed after putting your text content
\usepackage{graphicx}
\usepackage{natbib}
\usepackage{doi}
\usepackage{multirow}
\usepackage{amssymb}
\usepackage{algorithm,algorithmic}

\title{High-Accuracy Model Predictive Control with Inverse Hysteresis for High-Speed Trajectory Tracking of Piezoelectric Fast Steering Mirror}

\newif\ifuniqueAffiliation
\uniqueAffiliationtrue

\ifuniqueAffiliation % Standard variant of author block
\author{ \href{https://orcid.org/0000-0002-2471-1574}{\includegraphics[scale=0.06]{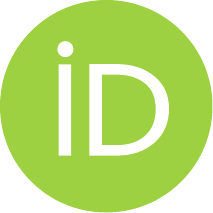}\hspace{1mm}Sen Yang} \\
	School of Astronautics and Aeronautics\\
	University of Electronic Science and Technology of China\\
	Chengdu, China 611731 \\
	\texttt{yang\_jansen@163.com} \\
	\And
	\href{https://orcid.org/0000-0000-0000-0000}{\includegraphics[scale=0.06]{orcid.pdf}\hspace{1mm}Xiaofeng Li} \\
	School of Astronautics and Aeronautics\\
	University of Electronic Science and Technology of China\\
	Chengdu, China 611731 \\
	\texttt{lxf3203433@uestc.edu.cn} \\
}
\else
\usepackage{authblk}

\newbox{\orcid}\sbox{\orcid}{\includegraphics[scale=0.06]{orcid.pdf}} 
\author[1]{%
	\href{https://orcid.org/0000-0002-2471-1574}{\usebox{\orcid}\hspace{1mm}Sen Yang\thanks{\texttt{yang\_jansen@163.com}}}%
}
\author[1]{%
	\href{https://orcid.org/0000-0000-0000-0000}{\usebox{\orcid}\hspace{1mm}Xiaofeng Li\thanks{\texttt{lxf3203433@uestc.edu.cn}}}%
}
\affil[1]{School of Astronautics and Aeronautics, University of Electronic Science and Technology of China, Chengdu, China 611731}
\fi

\renewcommand{\shorttitle}{\textit{arXiv} Template}

\begin{document}
\maketitle

\begin{abstract}
Piezoelectric fast steering mirrors (PFSM) are widely utilized in beam precision-pointing systems but encounter considerable challenges in achieving high-precision tracking of fast trajectories due to nonlinear hysteresis and mechanical dual-axis cross-coupling. This paper proposes a model predictive control (MPC) approach integrated with a hysteresis inverse based on the Hammerstein modeling structure of the PFSM. The MPC is designed to decouple the rate-dependent dual-axis linear components, with an augmented error integral variable introduced in the state space to eliminate steady-state errors. Moreover, proofs of zero steady-state error and disturbance rejection are provided. The hysteresis inverse model is then cascaded to compensate for the rate-independent nonlinear components. Finally, PFSM tracking experiments are conducted on step, sinusoidal, triangular, and composite trajectories. Compared to traditional model-free and existing model-based controllers, the proposed method significantly enhances tracking accuracy, demonstrating superior tracking performance and robustness to frequency variations. These results offer valuable insights for engineering applications.
\end{abstract}

% keywords can be removed

A fast steering mirror (FSM) is a widely used optical component that changes the direction of a beam by quickly and precisely adjusting its deflection angle. With its high control bandwidth, rapid response time, and capability to effectively suppress high-frequency disturbances, the FSM is integral to fine-tracking systems \cite{zhou2008design,haber2023data}, serving a critical role in both industrial and military applications.

The FSM primarily consists of a supporting structure, position sensors, actuators, and a controller. Supporting structures are typically categorized into two types. Compared to bearing-supported structures \cite{xu2012fast}, flexure hinge-supported structures \cite{zhao2020design} offer the significant advantage of eliminating frictional forces during movement, thereby avoiding sudden changes in force direction when the speed crosses zero, which reduces the anti-disturbance pressure. Actuators also come in two varieties. Compared to voice coil motor actuators \cite{wu2011large}, piezoelectric fast steering mirrors (PFSM) \cite{han2022design} provide higher control bandwidth, finer resolution (down to nanoradians), a more compact structure, and enhanced disturbance suppression capabilities. Consequently, PFSMs with flexure hinge-supported structures are more commonly used in beam precision-pointing systems , particularly in applications such as free-space optical communication \cite{kluk2007advanced,bekkali2022new} and laser precision machining \cite{csencsics2017system,zhong2022design}.

Although PFSM offers numerous advantages, it present challenges such as nonlinearity, multivariable interaction, and strong coupling. These issues manifest in the hysteresis nonlinearity and creep behavior between the input voltage and output displacement of the piezoelectric actuator (PEA), as well as in the mechanical dynamics and cross-coupling between the dual axes. To improve the control performance of PFSMs, researchers have proposed various control methods to mitigate these nonlinearities.

Early digital processors exhibited low computational performance and substantial latency, limiting their ability to achieve high control bandwidth. Consequently, analog circuits were employed. For instance, a compensator developed by MIT Lincoln Laboratory utilized digital switches to control analog voltage \cite{hawe2006control}. However, electrical control alone could not resolve the inherent mechanical issues, thereby hindering further increases in bandwidth.

With continuous technological advancements, particularly the development of high-performance digital signal processors and high-sampling-frequency analog-to-digital converters (ADC) and digital-to-analog converters (DAC), digital controllers have gradually supplanted traditional analog controllers. Ref. \cite{csencsics2017parametric} introduced a tuning method for proportional-integral-derivative (PID) controllers applicable to FSM, providing parameters that balance the robustness and performance of the closed-loop control system. In Ref. \cite{deng2017enhanced}, an improved disturbance observer was proposed and integrated with classical cascaded multi-loop feedback control, significantly enhancing the ability to suppress line-of-sight disturbances. These model-free control strategies treat nonlinearity as a disturbance, achieving good control performance at low frequencies. However, they fail to meet the high-precision tracking requirements during high-frequency movements.

Subsequently, scholars have explored model-based control methods to enhance the control bandwidth of PFSMs. In Ref. \cite{tang2010compensating}, the PFSM was modeled as a cascade of second-order mechanical resonance and first-order electronic hysteresis, with a direct inverse model (DIM) employed to mitigate nonlinearity in the feedforward path. Ref. \cite{wang2021laser} approximated the PFSM as a linear system and introduced an adaptive filtering variable step-size normalized least mean squares (FxVSNLMS) algorithm. Additionally, a saturated prescribed space-time adaptive sliding mode control (SPSASMC) was developed in work \cite{liu2023saturated} to address disturbances and input saturation, ensuring convergence within a specified time. These studies primarily focused on designing control algorithms for single-input-single-output (SISO) systems. 

However, the PFSM exhibits hysteresis and creep electrical coupling between dual PEAs within single axis, as well as mechanical cross-coupling between dual axes. The aforementioned methods, overlook these intrinsic properties and interrelationships, modeling the nonlinear multiple-input-multiple-output (MIMO) coupled system as dual separate SISO systems for the X and Y axes. This simplification limiting tracking performance. Moreover, PFSMs are constrained by input range limitations, necessitating the use of augmented saturation functions in most controllers, which results in a decline in closed-loop control performance \cite{wang2009model}. Consequently, there is an urgent need for a control method capable of handling MIMO model with constraints.

Model Predictive Control (MPC) offers a promising solution. MPC relies on the discrete-time model of the system, predicting its behavior over a future time horizon based on current outputs and anticipated inputs. It optimizes a predefined cost function to determine the optimal future control inputs, but only the first value of this control sequence is applied to the actual control system. This process is repeated at each sampling instance, hence the term receding horizon control \cite{rossiter2017model}. Due to its ability to handle MIMO dynamics and constraints, MPC has been extensively applied across various process control domains, including power electronics \cite{karamanakos2020model}, automotive applications \cite{norouzi2023integrating}, agriculture \cite{ding2018model}, renewable energy \cite{sultana2017review}, and high-precision tracking systems \cite{niu2020robust,huang2021robust}. Nevertheless, there are no reports of MPC aimed at enhancing the tracking performance of PFSM.

In this work, we first build upon our previous research \cite{yang2024first} to achieve high-precision modeling of the PFSM using a Hammerstein structure. Then propose a control method that involves a MPC to decouple the rate-dependent linear MIMO components and a cascaded hysteresis inversion model to compensate for the rate-independent nonlinearities. Classical MPC, which provides a solution based on the state and desired output, resembles proportional control and may exhibit steady-state errors. To mitigate this, we the error integration variables into the state space and provide proofs for zero steady-state error and disturbance rejection. Additionally, due to constraints on input ranges, a theoretical optimal solution cannot be directly derived. Instead, we utilize a coordinate descent algorithm to efficiently solve the constrained optimization, leveraging the convexity of its dual problem. We established a PFSM experimental platform and compared our proposed method with PID, DIM \cite{tang2010compensating}, FxVSNLMS \cite{wang2021laser}, and SPSASMC \cite{liu2023saturated}. The results demonstrate that our method offers superior tracking accuracy and robustness in handling frequency variations.

The remainder of this paper is organized as follows: In Section 2, we develop a high-precision model of the nonlinear MIMO-coupled PFSM system. Section 3 introduces a control method that combines MPC with hysteresis inversion. In Section 4, we experimentally evaluate the effectiveness of the proposed method. Finally, Section 5 concludes the paper.

\section{High-Precision Modeling of PFSM}
The PFSM is a nonlinear and complex system, with interconnected mechanical components that create a correlation between the dual-axis outputs. Building on our previous work \cite{yang2024first} adopting a Hammerstein structure to model the PFSM system, the nonlinear hysteresis is characterized by a modified Bouc-Wen model incorporating an asymmetric factor $\delta \dot{u}(t)u(t)$:
\begin{equation}
	\left\{ \begin{aligned}
		& {{v}_{h}}(t)=u(t)+h(t) \\ 
		& \dot{h}(t)=\alpha \dot{u}(t)-\beta \left| \dot{u}(t) \right|{{\left| h(t) \right|}^{n-1}}h(t)-\gamma \dot{u}(t){{\left| h(t) \right|}^{n}}+\delta \dot{u}(t)u(t) \\ 
	\end{aligned} \right.
	\label{eq:1}
\end{equation}
which preserves the rate-independent property of $h\left( t \right)$ while accurately describing the asymmetric physical characteristic of the PEA, which exhibits stronger resistance to deformation under higher driving voltage. Additionally, it enhances the ability to fit the nonlinear hysteresis behavior.

The rate-dependent dynamic characteristics are modeled using linear transfer functions that account for both creep dynamics ${{G}_{CRP}}$ and electromechanical dynamics ${{G}_{EM}}$.

Creep is a slow dynamic characteristic that describes the change in the voltage ${{v}_{C}}\left( t \right)$ across the equivalent capacitance of the PEA over time when the drive voltage $u\left( t \right)$ and the hysteresis voltage ${{v}_{h}}\left( t \right)$ remain constant. The creep effect is equivalently modeled using a spring-damper system \cite{gu2013motion}:
\begin{equation}
	{{G}_{CRP}}(s)=\frac{{{V}_{C}}(s)}{{{V}_{H}}(s)}=\prod\limits_{i}{\frac{s+{{z}_{i}}}{s+{{p}_{i}}}}
	\label{eq:2}
\end{equation}
where ${{V}_{C}}\left( s \right)$ is the Laplace transform of ${{v}_{C}}\left( t \right)$. The variable $i$ represents the order of the creep model, while ${{z}_{i}}$ and ${{p}_{i}}$ are the zeros and poles of the creep transfer function, respectively.

\begin{figure}[htpb]
	\centering\includegraphics[width=0.8\columnwidth]{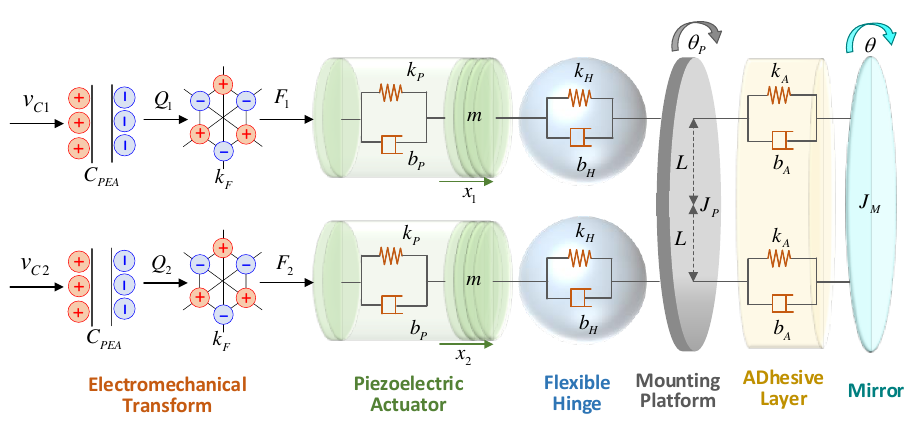}
	\caption{The electromechanical model of the PFSM.}
	\label{fig:1}
\end{figure}

The electromechanical model of the PFSM is illustrated in Fig. \ref{fig:1}. The charge $Q$ on the equivalent capacitance of the PEA is linearly converted into force $F$ [31], which then acts on the PEA and is transmitted to the mounting platform, resulting in the deflection of the mirror. By applying spring-damper models, a sixth-order transfer function for the electromechanical component is derived as follows \cite{yang2024first}:
\begin{equation}
	{{G}_{EM}}(s)=\frac{\theta (s)}{\Delta {{V}_{C}}(\text{s})}=\frac{{{b}_{3}}{{s}^{3}}+{{b}_{2}}{{s}^{2}}+{{b}_{1}}s+{{b}_{0}}}{{{a}_{6}}{{s}^{6}}+{{a}_{5}}{{s}^{5}}+{{a}_{4}}{{s}^{4}}+{{a}_{3}}{{s}^{3}}+{{a}_{2}}{{s}^{2}}+{{a}_{1}}s+{{a}_{0}}}
	\label{eq:3}
\end{equation}

The deflection of the PFSM exhibits two degrees of freedom, with coupling between the dual-axis outputs due to the mechanical interconnection of the system. Consequently, in addition to the single-axis transfer functions ${{G}_{EM,XX}}$ (X-axis to X-axis) and ${{G}_{EM,YY}}$, cross-axis transfer functions such as ${{G}_{EM,XY}}$ (X-axis to Y-axis) and ${{G}_{EM,YX}}$ are also present. By incorporating the creep dynamics Eq. (\ref{eq:2}), the dual-axis MIMO linear dynamic model is formulated as follows \cite{yang2024first}:
\begin{equation}
	\left[ \begin{matrix}
		{{\theta }_{X}}  \\
		{{\theta }_{Y}}  \\
	\end{matrix} \right]=\left[ \begin{matrix}
		{{G}_{CRP,X}}{{G}_{EM,XX}} & {{G}_{CRP,Y}}{{G}_{EM,YX}}  \\
		{{G}_{CRP,X}}{{G}_{EM,XY}} & {{G}_{CRP,Y}}{{G}_{EM,YY}}  \\
	\end{matrix} \right]\left[ \begin{matrix}
		\Delta {{V}_{H,X}}  \\
		\Delta {{V}_{H,Y}}  \\
	\end{matrix} \right]
	\label{eq:4}
\end{equation}

\section{MPC Combined with Inverse Bouc-Wen}
\begin{figure}[htpb]
	\centering\includegraphics[width=0.6\columnwidth]{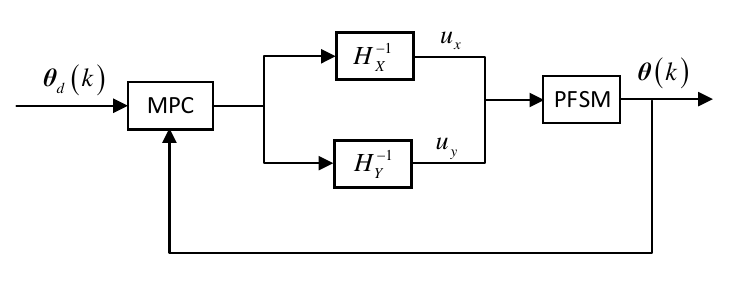}
	\caption{Block diagram of the MPC combined with inverse Bouc-Wen.}
	\label{fig:2}
\end{figure}

In this section, a MPC method combined with inverse Bouc-Wen is proposed, with the block diagram illustrated in Fig. \ref{fig:2}. The rate-dependent linear part is decoupled through MPC design, while the inverse Bouc-Wen models for the X-axis and Y-axis (denoted as $H_{X}^{-1}$ and $H_{Y}^{-1}$, respectively) are cascaded to compensate for nonlinear hysteresis. Consequently, the input voltages ${{u}_{x}}$ and ${{u}_{y}}$ for the dual axes are obtained, respectively.

According to Eq. (\ref{eq:4}), the inputs are the difference in hysteresis voltage between PEA1 and PEA2, given by  $\Delta {{v}_{h}}={{v}_{h1}}-{{v}_{h2}}$. Based on the proposed asymmetric hysteresis identification strategy, it holds that ${{v}_{h1}}+{{v}_{h2}}=100$, where 100 represents the fixed supply voltage applied to channel 3 of the PFSM. Defining ${{\tilde{v}}_{h1}}={\Delta {{v}_{h}}}/{2}$, then obtain:
\begin{equation}
	{{v}_{h1}}={{\tilde{v}}_{h1}}+50
	\label{eq:5}
\end{equation}

Thus, Eq. (\ref{eq:4}) is transformed into:
\begin{equation}
	\left[ \begin{matrix}
		{{\theta }_{X}}  \\
		{{\theta }_{Y}}  \\
	\end{matrix} \right]=\left[ \begin{matrix}
		2{{G}_{CRP,X}}{{G}_{EM,XX}} & 2{{G}_{CRP,Y}}{{G}_{EM,YX}}  \\
		2{{G}_{CRP,X}}{{G}_{EM,XY}} & 2{{G}_{CRP,Y}}{{G}_{EM,YY}}  \\
	\end{matrix} \right]\left[ \begin{matrix}
		{{{\tilde{V}}}_{H1,X}}  \\
		{{{\tilde{V}}}_{H1,Y}}  \\
	\end{matrix} \right]
	\label{eq:6}
\end{equation}

This continuous transfer function need be converted into a discrete transfer function using transformation such as Tustin method \cite{pfister2017discrete}, after which the discrete state-space equation is extracted:
\begin{equation}
	\left\{ \begin{aligned}
		& \mathbf{x}\left( k+1 \right)=A\mathbf{x}\left( k \right)+B{{{\tilde{\mathbf{v}}}}_{h}}\left( k \right) \\ 
		& \mathbf{\theta }\left( k \right)=C\mathbf{x}\left( k \right)+D{{\tilde{\mathbf{v}}}_{h}}\left( k \right) \\ 
	\end{aligned} \right.
	\label{eq:7}
\end{equation}
where ${{\mathbf{\tilde{v}}}_{h}}={{\left[ {{{\tilde{v}}}_{h1,x}},{{{\tilde{v}}}_{h1,y}} \right]}^{T}}\in {{\mathbb{R}}^{2\times 1}}$ and $\mathbf{\theta }={{\left[ {{\theta }_{x}},{{\theta }_{y}} \right]}^{T}}\in {{\mathbb{R}}^{2\times 1}}$ are the time-domain representations of ${{\left[ {{{\tilde{V}}}_{H1,X}},{{{\tilde{V}}}_{H1,Y}} \right]}^{T}}$ and ${{\left[ {{\theta }_{X}},{{\theta }_{Y}} \right]}^{T}}$, respectively. $\mathbf{x}\in {{\mathbb{R}}^{n\times 1}}$ is state vector, $A\in {{\mathbb{R}}^{n\times n}}$, $B\in {{\mathbb{R}}^{n\times 2}}$, $C\in {{\mathbb{R}}^{2\times n}}$, and $D\in {{\mathbb{R}}^{2\times 2}}$.

The integral of the tracking error is augmented into the state vector to eliminate steady-state errors. The error integral is defined as follows:
\begin{equation}
	\mathbf{h}\left( k+1 \right)=\mathbf{h}\left( k \right)+{{K}_{I}}\left[ {{\mathbf{\theta }}_{d}}\left( k+1 \right)-\mathbf{\theta }\left( k+1 \right) \right]
	\label{eq:8}
\end{equation}
where ${{K}_{I}}\in {{\mathbb{R}}^{2\times 2}}$ is the integral gain matrix, and ${{\mathbf{\theta }}_{d}}\left( k \right)={{\left[ {{\theta }_{x,d}}\left( k \right),{{\theta }_{y,d}}\left( k \right) \right]}^{T}}\in {{\mathbb{R}}^{2\times 1}}$ represents the reference vector at instant $k$. Substituting into Eq. (\ref{eq:7}) yields:
\begin{equation}
	\begin{aligned}
		& \mathbf{h}\left( k+1 \right)=\mathbf{h}\left( k \right)-{{K}_{I}}CA\mathbf{x}\left( k \right)-{{K}_{I}}CB{{{\tilde{\mathbf{v}}}}_{h}}\left( k \right) \\ 
		& \ \ \ \ \ \ \ \ \ \ \ \ \ \ \ -{{K}_{I}}D{{{\tilde{\mathbf{v}}}}_{h}}\left( k+1 \right)+{{K}_{I}}{{\mathbf{\theta }}_{d}}\left( k+1 \right) \\ 
	\end{aligned}
	\label{eq:9}
\end{equation}

Define the augmented system state vector:
\begin{equation}
	\mathbf{s}\left( k \right)=\left[ \begin{matrix}
		\mathbf{x}\left( k \right)  \\
		\mathbf{h}\left( k \right)  \\
	\end{matrix} \right]
	\label{eq:10}
\end{equation}

The modified state equation can be organized into matrix form as:
\begin{equation}
	\begin{aligned}
		& \left[ \begin{matrix}
			\mathbf{x}\left( k+1 \right)  \\
			\mathbf{h}\left( k+1 \right)  \\
		\end{matrix} \right]=\left[ \begin{matrix}
			A & 0  \\
			-{{K}_{I}}CA & I  \\
		\end{matrix} \right]\left[ \begin{matrix}
			\mathbf{x}\left( k \right)  \\
			\mathbf{h}\left( k \right)  \\
		\end{matrix} \right] \\ 
		& \ \ \ \ \ \ \ \ \ \ \ \ \ \ \ \ +\left[ \begin{matrix}
			B & 0  \\
			-{{K}_{I}}CB & -{{K}_{I}}D  \\
		\end{matrix} \right]\left[ \begin{matrix}
			{{{\tilde{\mathbf{v}}}}_{h}}\left( k \right)  \\
			{{{\tilde{\mathbf{v}}}}_{h}}\left( k+1 \right)  \\
		\end{matrix} \right]+\left[ \begin{matrix}
			0  \\
			{{K}_{I}}  \\
		\end{matrix} \right]{{\mathbf{\theta }}_{d}}\left( k+1 \right) \\ 
	\end{aligned}
	\label{eq:11}
\end{equation}
or simplified to:
\begin{equation}
	\mathbf{s}\left( k+1 \right)=F\mathbf{s}\left( k \right)+{{R}_{1}}{{\tilde{\mathbf{v}}}_{h}}\left( k \right)+{{R}_{2}}{{\tilde{\mathbf{v}}}_{h}}\left( k+1 \right)+Q{{\mathbf{\theta }}_{d}}\left( k+1 \right)
	\label{eq:12}
\end{equation}
where 
$F=\left[ \begin{matrix}
	A & 0  \\
	-{{K}_{I}}CA & I  \\
\end{matrix} \right]\in {{\mathbb{R}}^{\left( n+2 \right)\times \left( n+2 \right)}}$,
${{R}_{1}}=\left[ \begin{matrix}
	B  \\
	-{{K}_{I}}CB  \\
\end{matrix} \right]\in {{\mathbb{R}}^{\left( n+2 \right)\times 2}}$,
${{R}_{2}}=\left[ \begin{matrix}
	0  \\
	-{{K}_{I}}D  \\
\end{matrix} \right]\in {{\mathbb{R}}^{\left( n+2 \right)\times 2}}$,
$Q=\left[ \begin{matrix}
	0  \\
	{{K}_{I}}  \\
\end{matrix} \right]\in {{\mathbb{R}}^{\left( n+2 \right)\times 2}}$.

By combining the state-space equations from instant $k+1$ to $k+N$, the state equation set over the prediction horizon N is derived iteratively:
\begin{equation}
	\mathbf{S}=\Gamma \mathbf{s}\left( k \right)+\Upsilon {{\tilde{\mathbf{v}}}_{h}}\left( k \right)+\Phi {{\tilde{\mathbf{V}}}_{h}}+\Psi {{\mathbf{\Theta }}_{d}}
	\label{eq:13}
\end{equation}
where
$\mathbf{S}={{\left[ \begin{matrix}
			\mathbf{s}{{\left( k+1 \right)}^{T}} & \cdots  & \mathbf{s}{{\left( k+N \right)}^{T}}  \\
		\end{matrix} \right]}^{T}}$,
${{\tilde{\mathbf{V}}}_{h}}={{\left[ \begin{matrix}
			{{{\tilde{\mathbf{v}}}}_{h}}{{\left( k+1 \right)}^{T}} & \cdots  & {{{\tilde{\mathbf{v}}}}_{h}}{{\left( k+N \right)}^{T}}  \\
		\end{matrix} \right]}^{T}}$,
${{\mathbf{\Theta }}_{d}}={{\left[ \begin{matrix}
			{{\mathbf{\theta }}_{d}}{{\left( k+1 \right)}^{T}} & \cdots  & {{\mathbf{\theta }}_{d}}{{\left( k+N \right)}^{T}}  \\
		\end{matrix} \right]}^{T}}$,

$\Gamma =\left[ \begin{matrix}
	F  \\
	{{F}^{2}}  \\
	\vdots   \\
	{{F}^{N}}  \\
\end{matrix} \right]\in {{\mathbb{R}}^{N\left( n+2 \right)\times \left( n+2 \right)}}$,
$\Upsilon =\left[ \begin{matrix}
	{{R}_{1}}  \\
	F{{R}_{1}}  \\
	\vdots   \\
	{{F}^{N-1}}{{R}_{1}}  \\
\end{matrix} \right]\in {{\mathbb{R}}^{N\left( n+2 \right)\times 2}}$,

$\Phi =\left[ \begin{matrix}
	{{R}_{2}} & 0 & 0 & \cdots  & 0  \\
	F{{R}_{2}}+{{R}_{1}} & {{R}_{2}} & 0 & \cdots  & 0  \\
	{{F}^{2}}{{R}_{2}}+F{{R}_{1}} & F{{R}_{2}}+{{R}_{1}} & {{R}_{2}} & \ddots  & \vdots   \\
	\vdots  & \vdots  & \ddots  & \ddots  & 0  \\
	{{F}^{N-1}}{{R}_{2}}+{{F}^{N-2}}{{R}_{1}} & {{F}^{N-2}}{{R}_{2}}+{{F}^{N-3}}{{R}_{1}} & \cdots  & F{{R}_{2}}+{{R}_{1}} & {{R}_{2}}  \\
\end{matrix} \right]\in {{\mathbb{R}}^{N\left( n+2 \right)\times 2N}}$,

$\Psi =\left[ \begin{matrix}
	Q & 0 & \cdots  & 0  \\
	FQ & Q & \ddots  & \vdots   \\
	\vdots  & \vdots  & \ddots  & 0  \\
	{{F}^{N-1}}Q & {{F}^{N-2}}Q & \cdots  & Q  \\
\end{matrix} \right]\in {{\mathbb{R}}^{N\left( n+2 \right)\times 2N}}$.

Subsequently, based on the second equality in Eq. (\ref{eq:7}), the future N-step output vector $\mathbf{\Theta }={{\left[ \begin{matrix}
			\mathbf{\theta }{{\left( k+1 \right)}^{T}} & \cdots  & \mathbf{\theta }{{\left( k+N \right)}^{T}}  \\
		\end{matrix} \right]}^{T}}$ is given by:
\begin{equation}
	\mathbf{\Theta }=\tilde{C}\mathbf{S}+\tilde{D}{{\tilde{\mathbf{V}}}_{h}}
	\label{eq:14}
\end{equation}
where 
$\tilde{C}=\left[ \begin{matrix}
	C & 0 & {} & {} & {} & {} & {}  \\
	{} & {} & C & 0 & {} & {} & {}  \\
	{} & {} & {} & {} & \ddots  & {} & {}  \\
	{} & {} & {} & {} & {} & C & 0  \\
\end{matrix} \right]\in {{\mathbb{R}}^{2N\times N\left( n+2 \right)}}$,
$\tilde{D}=\left[ \begin{matrix}
	D & {} & {}  \\
	{} & \ddots  & {}  \\
	{} & {} & D  \\
\end{matrix} \right]\in {{\mathbb{R}}^{2N\times 2N}}$.

Additionally, the equation set for the future N-step error integral vector $\mathbf{H}={{\left[ \begin{matrix}
			\mathbf{h}{{\left( k+1 \right)}^{T}} & \cdots  & \mathbf{h}{{\left( k+N \right)}^{T}}  \\
		\end{matrix} \right]}^{T}}$ is:
\begin{equation}
	\mathbf{H}=\tilde{I}\mathbf{S}
	\label{eq:15}
\end{equation}
where
$\tilde{I}=\left[ \begin{matrix}
	0 & I & {} & {} & {} & {} & {}  \\
	{} & {} & 0 & I & {} & {} & {}  \\
	{} & {} & {} & {} & \ddots  & {} & {}  \\
	{} & {} & {} & {} & {} & 0 & I  \\
\end{matrix} \right]\in {{\mathbb{R}}^{2N\times N\left( n+2 \right)}}$.

Furthermore, the future N-step input increment vector $\Delta {{\tilde{\mathbf{V}}}_{h}}={{\left[ \begin{matrix}
			\Delta {{{\tilde{\mathbf{v}}}}_{h}}{{\left( k+1 \right)}^{T}} & \cdots  & \Delta {{{\tilde{\mathbf{v}}}}_{h}}{{\left( k+N \right)}^{T}}  \\
		\end{matrix} \right]}^{T}}$ can be derived from the decision vector ${{\tilde{\mathbf{V}}}_{h}}$ and the current input ${{\tilde{\mathbf{v}}}_{h}}\left( k \right)$, obtaining:
\begin{equation}
	\Delta {{\tilde{\mathbf{V}}}_{h}}=P{{\tilde{\mathbf{V}}}_{h}}-K{{\tilde{\mathbf{v}}}_{h}}\left( k \right)
	\label{eq:16}
\end{equation}
where
$P=\left[ \begin{matrix}
	I & {} & {} & {}  \\
	-I & I & {} & {}  \\
	{} & \ddots  & \ddots  & {}  \\
	{} & {} & -I & I  \\
\end{matrix} \right]\in {{\mathbb{R}}^{2N\times 2N}}$,
$K=\left[ \begin{matrix}
	I  \\
	0  \\
	\vdots   \\
	0  \\
\end{matrix} \right]\in {{\mathbb{R}}^{2N\times 2}}$.

We aim to minimize the difference between the predicted and the desired outputs, as well as the change in the control variables, which indicates achieving control stability. To obtain the optimal control input vector ${{\tilde{\mathbf{V}}}_{h}}$, the cost function $J$ is defined as:
\begin{equation}
	\begin{aligned}
		& J=\frac{1}{2}\left\| {{\mathbf{\Theta }}_{d}}-\mathbf{\Theta } \right\|_{2}^{2}+\frac{1}{2}\left\| \mathbf{H} \right\|_{2}^{2}+\frac{\rho }{2}\left\| \Delta {{{\tilde{\mathbf{V}}}}_{h}} \right\|_{2}^{2} \\ 
		& \ \ \ =\frac{1}{2}\tilde{\mathbf{V}}_{h}^{T}M{{{\tilde{\mathbf{V}}}}_{h}}+{{\mathbf{m}}^{T}}{{{\tilde{\mathbf{V}}}}_{h}}+{{m}_{0}} \\ 
	\end{aligned}
	\label{eq:17}
\end{equation}
where $\rho $ is the penalty coefficient for the control increment vector. The error integral vector can be adjusted through ${{K}_{I}}$, for which the corresponding penalty coefficient is set to 1. The other parameters are as follows:
\begin{equation}
	\begin{aligned}
		& M={{\left( \tilde{C}\Phi +\tilde{D} \right)}^{T}}\left( \tilde{C}\Phi +\tilde{D} \right)+{{\Phi }^{T}}{{{\tilde{I}}}^{T}}\tilde{I}\Phi +\rho {{P}^{T}}P\in {{\mathbb{R}}^{2N\times 2N}},\ \  \\ 
		& \mathbf{m}={{\left( \tilde{C}\Phi +\tilde{D} \right)}^{T}}\mathbf{\beta }+{{\Phi }^{T}}{{{\tilde{I}}}^{T}}\tilde{I}\mathbf{\eta }-\rho {{P}^{T}}K{{{\tilde{\mathbf{v}}}}_{h}}\left( k \right)\in {{\mathbb{R}}^{2N\times 1}}, \\ 
		& {{m}_{0}}=\frac{1}{2}{{\mathbf{\beta }}^{T}}\mathbf{\beta }+\frac{1}{2}{{\mathbf{\eta }}^{T}}{{{\tilde{I}}}^{T}}\tilde{I}\mathbf{\eta }+\frac{\rho }{2}\tilde{\mathbf{v}}_{h}^{T}\left( k \right){{K}^{T}}K{{{\tilde{\mathbf{v}}}}_{h}}\left( k \right)\in \mathbb{R},\  \\ 
		& \mathbf{\eta }=\Gamma \mathbf{s}\left( k \right)+\Upsilon {{{\tilde{\mathbf{v}}}}_{h}}\left( k \right)+\Psi {{\mathbf{\Theta }}_{d}}\in {{\mathbb{R}}^{N\left( n+2 \right)\times 1}}, \\ 
		& \mathbf{\beta }=\tilde{C}\mathbf{\eta }-{{\mathbf{\Theta }}_{d}}\in {{\mathbb{R}}^{2N\times 1}} \\ 
	\end{aligned}
	\label{eq:18}
\end{equation}
where ${{m}_{0}}$ is a constant term independent of ${{\tilde{\mathbf{V}}}_{h}}$, and can be ignored in the optimization of the cost function. Considering the constraints on the ranges of input signals and output angles, the problem formulated in Eq. (\ref{eq:17}) is transformed into the following quadratic programming (QP) problem:
\begin{equation}
	\tilde{\mathbf{V}}_{h}^{*}=\underset{{{{\tilde{\mathbf{V}}}}_{h}}}{\mathop{\arg \min }}\,\frac{1}{2}\tilde{\mathbf{V}}_{h}^{T}M{{\tilde{\mathbf{V}}}_{h}}+{{\mathbf{m}}^{T}}{{\tilde{\mathbf{V}}}_{h}},\ \ \text{s}.\text{t}.\ \ W{{\tilde{\mathbf{V}}}_{h}}\le \mathbf{b}
	\label{eq:19}
\end{equation}
where 

$W=\left[ \begin{matrix}
	\tilde{C}\Phi +\tilde{D}  \\
	-\tilde{C}\Phi -\tilde{D}  \\
	I  \\
	-I  \\
\end{matrix} \right]\in {{\mathbb{R}}^{8N\times 2N}}$,
$\mathbf{b}=\left[ \begin{matrix}
	{{I}_{\theta }}{{\mathbf{\theta }}_{\max }}-\tilde{C}\mathbf{\eta }  \\
	-{{I}_{\theta }}{{\mathbf{\theta }}_{\min }}+\tilde{C}\mathbf{\eta }  \\
	{{I}_{v}}{{{\tilde{\mathbf{v}}}}_{h,\max }}  \\
	-{{I}_{v}}{{{\tilde{\mathbf{v}}}}_{h,\min }}  \\
\end{matrix} \right]\in {{\mathbb{R}}^{8N\times 1}}$,
${{I}_{\theta }}=\left[ \begin{matrix}
	I  \\
	I  \\
	\vdots   \\
	I  \\
\end{matrix} \right]\in {{\mathbb{R}}^{2N\times 2}}$,
${{I}_{v}}=\left[ \begin{matrix}
	I  \\
	I  \\
	\vdots   \\
	I  \\
\end{matrix} \right]\in {{\mathbb{R}}^{2N\times 2}}$,

${{\mathbf{\theta }}_{\min }}\in {{\mathbb{R}}^{2\times 1}}$ and ${{\mathbf{\theta }}_{\max }}\in {{\mathbb{R}}^{2\times 1}}$ represent the lower and upper bounds of the output angle range, respectively. ${{\tilde{\mathbf{v}}}_{h,\min }}\in {{\mathbb{R}}^{2\times 1}}$ and ${{\tilde{\mathbf{v}}}_{h,\max }}\in {{\mathbb{R}}^{2\times 1}}$ represent the lower and upper bounds of the input signal range, respectively.

Despite the availability of various optimization solvers, such as Hildreth \cite{wang2009model} and the open-source framework acado \cite{verschueren2022acados}, solving the constrained QP remains challenging within limited sampling times on resource-constrained hardware. To address this, constraints can be incorporated into the cost function to form the Lagrange function as follows:
\begin{equation}
	L\left( {{{\tilde{\mathbf{V}}}}_{h}},\mathbf{\mu } \right)=\frac{1}{2}\tilde{\mathbf{V}}_{h}^{T}M{{\tilde{\mathbf{V}}}_{h}}+{{\mathbf{m}}^{T}}{{\tilde{\mathbf{V}}}_{h}}+{{\mathbf{\mu }}^{T}}\left( W{{{\tilde{\mathbf{V}}}}_{h}}-\mathbf{b} \right)
	\label{eq:20}
\end{equation}
where the Lagrange multiplier vector $\mathbf{\mu }\in {{\mathbb{R}}^{8N\times 1}}$ is used to penalize the inequality constraints. The minimum of the function (\ref{eq:20}) is reached when ${\partial L}/{\partial {{{\tilde{\mathbf{V}}}}_{h}}=0}$, yielding:
\begin{equation}
	{{\tilde{\mathbf{V}}}_{h}}=-{{M}^{-1}}\left( \mathbf{m}+{{W}^{T}}\mathbf{\mu } \right)
	\label{eq:21}
\end{equation}

By substituting Eq. (\ref{eq:21}) into Eq. (\ref{eq:20}), the dual problem corresponding to Eq. (\ref{eq:19}) is derived as:
\begin{equation}
	{{\mathbf{\mu }}^{*}}\triangleq \underset{\mathbf{\mu }}{\mathop{\arg \max }}\,\left( -\frac{1}{2}{{\mathbf{\mu }}^{T}}G\mathbf{\mu }-{{\mathbf{g}}^{T}}\mathbf{\mu }-{{g}_{0}} \right)\text{,}\ \ \text{s}.\text{t}.\ \ \mathbf{\mu }\ge 0
	\label{eq:22}
\end{equation}
where
\begin{equation}
	G=W{{M}^{-1}}{{W}^{T}}\in {{\mathbb{R}}^{8N\times 8N}},\ \mathbf{g}=W{{M}^{-1}}\mathbf{m}+\mathbf{b}\in {{\mathbb{R}}^{8N\times 1}},\ {{g}_{0}}=\frac{1}{2}{{\mathbf{m}}^{T}}{{M}^{-1}}\mathbf{m}\in \mathbb{R}
	\label{eq:23}
\end{equation}

Since ${{g}_{0}}$ is independent of $\mathbf{\mu }$, it can be disregarded during optimization. The dual problem (\ref{eq:22}) is then reformulated as:
\begin{equation}
	{{\mathbf{\mu }}^{*}}\triangleq \underset{\mathbf{\mu }}{\mathop{\arg \min }}\,\frac{1}{2}{{\mathbf{\mu }}^{T}}G\mathbf{\mu }+{{\mathbf{g}}^{T}}\mathbf{\mu }\text{,}\ \ \text{s}.\text{t}.\ \ \mathbf{\mu }\ge 0
	\label{eq:24}
\end{equation}

Compared to solving the primal problem (\ref{eq:19}) directly, the dual problem (\ref{eq:24}) presents a simpler constraint, $\mathbf{\mu }\ge 0$, which enhances computational efficiency. In practical PFSM control, constraining only the input range ensures that the output remains within the desired bounds. Consequently, given that both $M$ and $G$ are symmetric positive definite as established in Eqs. (\ref{eq:18}) and (\ref{eq:23}), the function (\ref{eq:24}) is strictly convex with respect to the elements of $\mathbf{\mu }$. This allows for the application of the coordinate descent method \cite{wright2015coordinate, dong2023embedded} to solve it iteratively.

Let ${{\mu }_{i}}$ denote the $i$-th element of the vector $\mathbf{\mu }$. By setting the partial derivative of Eq. (\ref{eq:24}) with respect to ${{\mu }_{i}}$ equal to zero, the minimum of the dual optimization problem along the direction of ${{\mu }_{i}}$ is attained at ${{\hat{\mu }}_{i}}$, which is:
\begin{equation}
	{{\hat{\mu }}_{i}}=-\frac{1}{{{G}_{ii}}}\left( \sum\limits_{j\ne i}{{{G}_{ij}}{{\mu }_{j}}}+{{\mathbf{g}}_{i}} \right)={{\mu }_{i}}-\frac{1}{{{G}_{ii}}}\left( {{G}_{i:}}\mathbf{\mu }+{{\mathbf{g}}_{i}} \right)
	\label{eq:25}
\end{equation}
where ${{\mathbf{g}}_{i}}$ represent the $i$-th element of the vector $\mathbf{g}$, ${{G}_{ij}}$ represent the element in the $i$-th row and $j$-th column of the matrix $G$, and ${{G}_{i:}}$ is the $i$-th row of matrix $G$. Combining the constraint ${{\mu }_{i}}\ge 0$, the update formula for ${{\mu }_{i}}$ using the coordinate descent method is given by:
\begin{equation}
	\mu _{i}^{*}=\max \left\{ {{{\hat{\mu }}}_{i}},\ 0 \right\}
	\label{eq:26}
\end{equation}

By iteratively performing Eq. (\ref{eq:26}), the optimal solution ${{\mathbf{\mu }}^{*}}$ can be obtained. From Eq. (\ref{eq:21}), it follows that the optimal solution $\mathbf{\tilde{V}}_{h}^{*}$ for the primal problem satisfies:
\begin{equation}
	\tilde{\mathbf{V}}_{h}^{*}=-{{M}^{-1}}\left( \mathbf{m}+{{W}^{T}}{{\mathbf{\mu }}^{*}} \right)
	\label{eq:27}
\end{equation}

Since the input is two-dimension, let the first two elements represent the control signals at instant $k+1$, then:
\begin{equation}
	{{\tilde{\mathbf{v}}}_{h}}\left( k+1 \right)=\left[ \begin{matrix}
		I & 0 & \cdots  & 0  \\
	\end{matrix} \right]\tilde{\mathbf{V}}_{h}^{*}=-M_{I}^{-1}\left( \mathbf{m}+{{W}^{T}}{{\mathbf{\mu }}^{*}} \right)
	\label{eq:28}
\end{equation}
where $M_{I}^{-1}$ is the first two rows of the matrix ${{M}^{-1}}$. The complete solution process is summarized in Algorithm \ref{alg:1}. For given parameters $N$ and $\rho $, the matrices $\Gamma $, $\Upsilon $, $\Phi $, $\Psi $, $\tilde{C}$, $\tilde{D}$, $\tilde{I}$, $P$, $K$, $W$, $M$, and $G$ are time-invariant, enabling them to be precomputed, which reduces control delay. Additionally, as the optimal solution ${{\mathbf{\mu }}^{*}}$ of the dual problem evolves continuously over time, we can initialize $\mathbf{\mu }\left( k+1 \right)$ with the current step ${{\mathbf{\mu }}^{*}}\left( k \right)$ to accelerate the convergence of the algorithm.

\begin{algorithm}[H]  
	\renewcommand{\algorithmicrequire}{\textbf{Input:}}
	\renewcommand{\algorithmicensure}{\textbf{Output:}}
	\caption{The coordinate descent method for QP problem (\ref{eq:19}).}
	\label{alg:1}
	\begin{algorithmic}[1] % 控制是否有序号
		\REQUIRE  The current state $\mathbf{s}\left( k \right)$, the current control input ${{\tilde{\mathbf{v}}}_{h}}\left( k \right)$, the reference input vector ${{\mathbf{\Theta }}_{d}}$ from instant $k+1$ to $k+N$, and the current dual solution ${{\mathbf{\mu }}^{*}}\left( k \right)$, with ${{\mathbf{\mu }}^{*}}\left( 1 \right)={{0}_{8N\times 1}}$; % input 的内容
		\ENSURE The control input ${{\mathbf{\tilde{v}}}_{h}}\left( k+1 \right)$ and the dual solution ${{\mathbf{\mu }}^{*}}\left( k+1 \right)$; % output 的内容

		\STATE Let $\hat{\mathbf{\mu }}={{\mathbf{\mu }}^{*}}\left( k \right)$
		\STATE Calculate $\mathbf{m}$ and $\mathbf{b}$ of the original problem (\ref{eq:19});
		\STATE Calculate $\mathbf{g}$ of the dual problem (\ref{eq:24});
		
		% for loop
		\FORALL {$i=1,2,\cdots ,8N$}
		\STATE ${{\hat{\mu }}_{i}}=\max \left\{ {{{\hat{\mu }}}_{i}}-{\left( {{G}_{i:}}\hat{\mathbf{\mu }}+{{\mathbf{g}}_{i}} \right)}/{{{G}_{ii}}},\ 0 \right\}$;
		\ENDFOR
		
		\STATE \textbf{return} ${{\tilde{\mathbf{v}}}_{h}}\left( k+1 \right)=-M_{I}^{-1}\left( \mathbf{m}+{{W}^{T}}\hat{\mathbf{\mu }} \right)$, ${{\mathbf{\mu }}^{*}}\left( k+1 \right)=\hat{\mathbf{\mu }}$.
	\end{algorithmic}
\end{algorithm}

By obtaining the control input ${{\mathbf{\tilde{v}}}_{h}}\left( k+1 \right)$ through MPC, the voltage vector affected by hysteresis in dual axes can be determined using Eq. (\ref{eq:5}) as ${{\mathbf{v}}_{h}}\left( k+1 \right)={{\mathbf{\tilde{v}}}_{h}}\left( k+1 \right)+50$, where ${{\mathbf{v}}_{h}}={{\left[ {{v}_{h1,x}},{{v}_{h1,y}} \right]}^{T}}\in {{\mathbb{R}}^{2\times 1}}$. This process achieves the decoupling of the dual-axis mechanical cross, thereby enabling the subsequent cascading of the inverse Bouc-Wen model for each axis to compensate for nonlinear hysteresis, ultimately resulting in the input voltage for the dual axes. The discrete form of the inverse Bouc-Wen model for X-axis is expressed as follows:
\begin{equation}
	\left\{ \begin{aligned}
		& {{u}_{x}}\left( k \right)={{v}_{h1,x}}\left( k \right)-h\left( k \right) \\ 
		& \Delta h\left( k+1 \right)=\alpha \Delta {{u}_{x}}\left( k \right)-\beta \left| \Delta {{u}_{x}}\left( k \right) \right|{{\left| h\left( k \right) \right|}^{n-1}}h\left( k \right) \\ 
		& \ \ \ \ \ \ \ \ \ \ \ \ \ \ -\gamma \Delta {{u}_{x}}\left( k \right){{\left| h\left( k \right) \right|}^{n}}+\delta \Delta {{u}_{x}}\left( k \right){{u}_{x}}\left( k \right) \\ 
	\end{aligned} \right.
	\label{eq:29}
\end{equation}
where ${{u}_{x}}\left( k \right),{{v}_{h1,x}}\left( k \right)$, and $h\left( k \right)$ are all scalars known at instant $k$. By utilizing the second equation, $\Delta h\left( k+1 \right)$ can be computed, followed by the calculation of $h\left( k+1 \right)$. While ${{v}_{h1,x}}\left( k+1 \right)$ can be determined using the output ${{v}_{h1,x}}\left( k+1 \right)$ from the MPC. Subsequently, ${{u}_{x}}\left( k+1 \right)$ can be explicitly computed using the first equation. By iterating this process, the control voltage for the X-axis at each subsequent time step can be obtained. The same methodology is applicable to the Y-axis.

\subsection{Stability Analysis}
In this section, the characteristics of the proposed control scheme, specifically its zero steady-state error and disturbance rejection capabilities, are investigated. Eq. (\ref{eq:28}) can be rearranged further as follows:
\begin{equation}
	{{\tilde{\mathbf{v}}}_{h}}\left( k+1 \right)=-\hat{\Gamma }\mathbf{s}\left( k \right)-\hat{\Upsilon }{{\tilde{\mathbf{v}}}_{h}}\left( k \right)-\hat{\Psi }{{\mathbf{\Theta }}_{d}}-\hat{W}{{\mathbf{\mu }}^{*}}
	\label{eq:30}
\end{equation}
where
$\hat{\Gamma }=M_{I}^{-1}\left[ {{\left( \tilde{C}\Phi +\tilde{D} \right)}^{T}}\tilde{C}+{{\Phi }^{T}}{{{\tilde{I}}}^{T}}\tilde{I} \right]\Gamma \in {{\mathbb{R}}^{2\times \left( n+2 \right)}}$,

$\hat{\Upsilon }=M_{I}^{-1}\left[ {{\left( \tilde{C}\Phi +\tilde{D} \right)}^{T}}\tilde{C}\Upsilon +{{\Phi }^{T}}{{{\tilde{I}}}^{T}}\tilde{I}\Upsilon -\rho {{P}^{T}}K \right]\in {{\mathbb{R}}^{2\times 2}}$,

$\hat{\Psi }=M_{I}^{-1}\left[ {{\left( \tilde{C}\Phi +\tilde{D} \right)}^{T}}\left( \tilde{C}\Psi -I \right)+{{\Phi }^{T}}{{{\tilde{I}}}^{T}}\tilde{I}\Psi  \right]\in {{\mathbb{R}}^{2\times 2N}}$,
$\hat{W}=M_{I}^{-1}{{W}^{T}}\in {{\mathbb{R}}^{2\times 8N}}$.

Let $\hat{\Gamma }=[ \begin{matrix}
	{{\mathbf{f}}_{x}} & {{\mathbf{f}}_{h}}  \\
\end{matrix} ]$, $\hat{\Psi }=[ \begin{matrix}
{{\mathbf{\psi }}_{1}} & {{\mathbf{\psi }}_{2}} & \cdots  & {{\mathbf{\psi }}_{N}}  \\
\end{matrix} ]$, and $\hat{W}{{\mathbf{\mu }}^{*}}={{\mathbf{\omega }}_{0}}$. The state observer can be considered a linear combination of historical inputs and outputs \cite{cao2012inversion}:
\begin{equation}
	\begin{aligned}
		& {{\mathbf{f}}_{h}}\mathbf{x}\left( k \right)={{\mathbf{\tau }}_{1}}\mathbf{\theta }\left( k-1 \right)+{{\mathbf{\tau }}_{2}}\mathbf{\theta }\left( k-2 \right)+\cdots +{{\mathbf{\tau }}_{p}}\mathbf{\theta }\left( k-p \right) \\ 
		& \ \ \ \ \ \ \ \ \ \ \ +{{\mathbf{\kappa }}_{1}}{{{\tilde{\mathbf{v}}}}_{h}}\left( k-1 \right)+{{\mathbf{\kappa }}_{2}}{{{\tilde{\mathbf{v}}}}_{h}}\left( k-2 \right)+\cdots +{{\mathbf{\kappa }}_{q}}{{{\tilde{\mathbf{v}}}}_{h}}\left( k-q \right) \\ 
	\end{aligned}
	\label{eq:31}
\end{equation}

Substituting into Eq. (\ref{eq:30}) yields:
\begin{equation}
	\begin{aligned}
		& {{{\tilde{\mathbf{v}}}}_{h}}\left( k+1 \right)=-{{\mathbf{\tau }}_{1}}\mathbf{\theta }\left( k-1 \right)-{{\mathbf{\tau }}_{2}}\mathbf{\theta }\left( k-2 \right)-\cdots -{{\mathbf{\tau }}_{p}}\mathbf{\theta }\left( k-p \right) \\ 
		& \ \ \ \ \ \ \ \ \ \ \ \ \ \ \ \ -\hat{\Upsilon }{{{\tilde{\mathbf{v}}}}_{h}}\left( k \right)-{{\mathbf{\kappa }}_{1}}{{{\tilde{\mathbf{v}}}}_{h}}\left( k-1 \right)-{{\mathbf{\kappa }}_{2}}{{{\tilde{\mathbf{v}}}}_{h}}\left( k-2 \right)-\cdots -{{\mathbf{\kappa }}_{q}}{{{\tilde{\mathbf{v}}}}_{h}}\left( k-q \right)-{{\mathbf{f}}_{h}}\mathbf{h}\left( k \right) \\ 
		& \ \ \ \ \ \ \ \ \ \ \ \ \ \ \ \ -{{\mathbf{\psi }}_{1}}{{\mathbf{\theta }}_{d}}\left( k+1 \right)-{{\mathbf{\psi }}_{2}}{{\mathbf{\theta }}_{d}}\left( k+2 \right)-\cdots -{{\mathbf{\psi }}_{N}}{{\mathbf{\theta }}_{d}}\left( k+N \right)-{{\mathbf{\omega }}_{0}} \\ 
	\end{aligned}
	\label{eq:32}
\end{equation}

Then take the $\mathbf{z}$ transform:
\begin{equation}
	\begin{aligned}
		& {{{\tilde{\mathbf{v}}}}_{h}}\left( k+1 \right)=-\left( {{\mathbf{\tau }}_{1}}{{\mathbf{z}}^{-1}}+{{\mathbf{\tau }}_{2}}{{\mathbf{z}}^{-2}}+\cdots +{{\mathbf{\tau }}_{p}}{{\mathbf{z}}^{-p}} \right)\mathbf{\theta }\left( k \right)-\frac{{{\mathbf{f}}_{h}}\mathbf{z}{{K}_{I}}}{\mathbf{z}-1}\left[ {{\mathbf{\theta }}_{d}}\left( k \right)-\mathbf{\theta }\left( k \right) \right] \\ 
		& \ \ \ \ \ \ \ \ \ \ \ \ \ \ \ \ -\left( \hat{\Upsilon }+{{\mathbf{\kappa }}_{1}}{{\mathbf{z}}^{-1}}+{{\mathbf{\kappa }}_{2}}{{\mathbf{z}}^{-2}}+\cdots +{{\mathbf{\kappa }}_{q}}{{\mathbf{z}}^{-q}} \right){{{\tilde{\mathbf{v}}}}_{h}}\left( k \right) \\ 
		& \ \ \ \ \ \ \ \ \ \ \ \ \ \ \ \ -\left( \ {{\mathbf{\psi }}_{1}}{{\mathbf{z}}^{1}}+{{\mathbf{\psi }}_{2}}{{\mathbf{z}}^{2}}+\cdots +{{\mathbf{\psi }}_{N}}{{\mathbf{z}}^{N}} \right){{\mathbf{\theta }}_{d}}\left( k \right)-{{\mathbf{\omega }}_{0}} \\ 
	\end{aligned}
	\label{eq:33}
\end{equation}
where $\mathbf{z}$ is the forward shift operator. Further rearranging gives:
\begin{equation}
	{{L}_{v}}\left( \mathbf{z} \right){{\tilde{\mathbf{v}}}_{h}}\left( k \right)={{L}_{\theta }}\left( \mathbf{z} \right)\mathbf{\theta }\left( k \right)+{{L}_{d}}\left( \mathbf{z} \right){{\mathbf{\theta }}_{d}}\left( k \right)+\frac{{{L}_{h}}\left( \mathbf{z} \right)}{\mathbf{z}-1}\left[ {{\mathbf{\theta }}_{d}}\left( k \right)-\mathbf{\theta }\left( k \right) \right]-{{\mathbf{\omega }}_{0}}
	\label{eq:34}
\end{equation}
where 

${{L}_{v}}\left( \mathbf{z} \right)=\mathbf{z}+\hat{\Upsilon }+{{\mathbf{\kappa }}_{1}}{{\mathbf{z}}^{-1}}+{{\mathbf{\kappa }}_{2}}{{\mathbf{z}}^{-2}}+\cdots +{{\mathbf{\kappa }}_{q}}{{\mathbf{z}}^{-q}}$,
${{L}_{\theta }}\left( \mathbf{z} \right)=-{{\mathbf{\tau }}_{1}}{{\mathbf{z}}^{-1}}-{{\mathbf{\tau }}_{2}}{{\mathbf{z}}^{-2}}-\cdots -{{\mathbf{\tau }}_{p}}{{\mathbf{z}}^{-p}}$,

${{L}_{d}}\left( \mathbf{z} \right)=-{{\mathbf{\psi }}_{1}}{{\mathbf{z}}^{1}}-{{\mathbf{\psi }}_{2}}{{\mathbf{z}}^{2}}-\cdots -{{\mathbf{\psi }}_{N}}{{\mathbf{z}}^{N}}$,
${{L}_{h}}\left( \mathbf{z} \right)=-{{\mathbf{f}}_{h}}\mathbf{z}{{K}_{I}}$.

\begin{figure}[htpb]
	\centering\includegraphics[width=0.8\columnwidth]{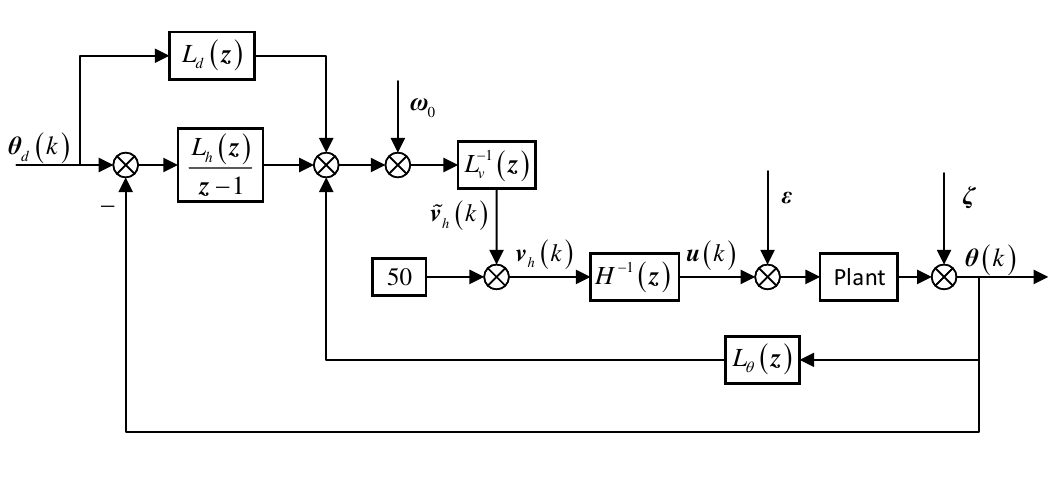}
	\caption{The block diagram of MPC.}
	\label{fig:3}
\end{figure}

Fig. \ref{fig:3} illustrates the block diagram of Eq. (\ref{eq:34}), where $\mathbf{\varepsilon }$ and $\mathbf{\zeta }$ are the input and output disturbances, respectively. The transfer function of system is $G\left( \mathbf{z} \right)={B\left( \mathbf{z} \right)}/{A\left( \mathbf{z} \right)}\;$. Therefore, the closed-loop transfer function between the output and the reference signal is expressed by:
\begin{equation}
	{{T}_{\theta }}\left( \mathbf{z} \right)=\frac{\mathbf{\theta }\left( \mathbf{z} \right)}{{{\mathbf{\theta }}_{d}}\left( \mathbf{z} \right)}=\frac{{{L}_{h}}\left( \mathbf{z} \right)B\left( \mathbf{z} \right)+\left( 1-{{\mathbf{z}}^{-1}} \right){{L}_{d}}\left( \mathbf{z} \right)B\left( \mathbf{z} \right)}{{{L}_{h}}\left( \mathbf{z} \right)B\left( \mathbf{z} \right)-\left( 1-{{\mathbf{z}}^{-1}} \right)\left[ {{L}_{v}}\left( \mathbf{z} \right)H\left( \mathbf{z} \right)A\left( \mathbf{z} \right)-{{L}_{\theta }}\left( \mathbf{z} \right)B\left( \mathbf{z} \right) \right]}
	\label{eq:35}
\end{equation}

There is $\underset{\mathbf{z}\to I}{\mathop{\lim }}\,{{T}_{\theta }}\left( \mathbf{z} \right)=I$ under the condition of closed-loop control stability, indicating that the steady-state error in tracking the reference input is zero. Additionally, the transfer functions from the disturbances ($\mathbf{\varepsilon }$ and $\mathbf{\zeta }$) to the output signal are given by:
\begin{equation}
	{{T}_{\varepsilon }}\left( \mathbf{z} \right)=\frac{\mathbf{\theta }\left( \mathbf{z} \right)}{\mathbf{\varepsilon }\left( \mathbf{z} \right)}=\frac{\left( 1-{{\mathbf{z}}^{-1}} \right){{L}_{v}}\left( \mathbf{z} \right)H\left( \mathbf{z} \right)B\left( \mathbf{z} \right)}{{{L}_{h}}\left( \mathbf{z} \right)B\left( \mathbf{z} \right)-\left( 1-{{\mathbf{z}}^{-1}} \right)\left[ {{L}_{v}}\left( \mathbf{z} \right)H\left( \mathbf{z} \right)A\left( \mathbf{z} \right)-{{L}_{\theta }}\left( \mathbf{z} \right)B\left( \mathbf{z} \right) \right]}
	\label{eq:36}
\end{equation}

\begin{equation}
	{{T}_{\zeta }}\left( \mathbf{z} \right)=\frac{\mathbf{\theta }\left( \mathbf{z} \right)}{\mathbf{\zeta }\left( \mathbf{z} \right)}=\frac{\left( 1-{{\mathbf{z}}^{-1}} \right){{L}_{v}}\left( \mathbf{z} \right)H\left( \mathbf{z} \right)A\left( \mathbf{z} \right)}{{{L}_{h}}\left( \mathbf{z} \right)B\left( \mathbf{z} \right)-\left( 1-{{\mathbf{z}}^{-1}} \right)\left[ {{L}_{v}}\left( \mathbf{z} \right)H\left( \mathbf{z} \right)A\left( \mathbf{z} \right)-{{L}_{\theta }}\left( \mathbf{z} \right)B\left( \mathbf{z} \right) \right]}
	\label{eq:37}
\end{equation}

There is $\underset{\mathbf{z}\to I}{\mathop{\lim }}\,{{T}_{\varepsilon }}\left( \mathbf{z} \right)=0$ and $\underset{\mathbf{z}\to I}{\mathop{\lim }}\,{{T}_{\zeta }}\left( \mathbf{z} \right)=0$ under the condition of closed-loop control stability, indicating that the control system can effectively reject input and output disturbances.

\section{Experiment}
\begin{figure}[htpb]
	\centering\includegraphics[width=0.7\columnwidth]{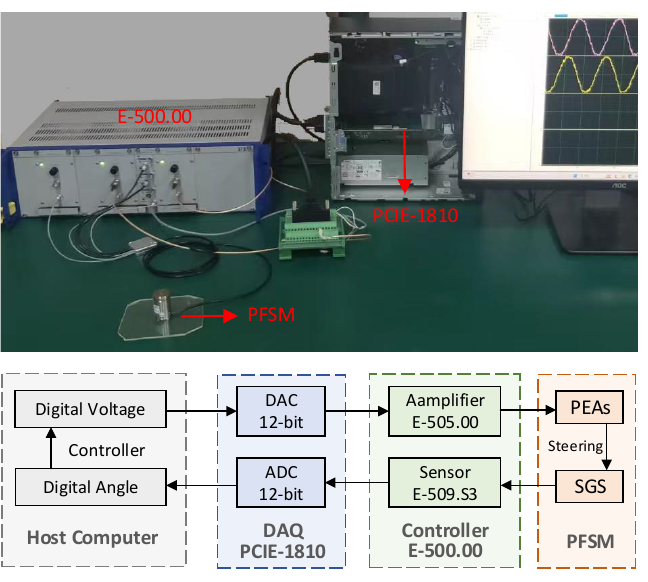}
	\caption{The setup of the experimental platform.}
	\label{fig:4}
\end{figure}

An experimental platform for the PFSM system is constructed, as shown in Fig. \ref{fig:4}. It consists of a modular piezoelectric controller (PI, E-500.00), a dual-axis PFSM (PI, S-330.2SL) equipped with strain gauge sensor (SGS), a data acquisition card (AdvanTech, PCIE-1810), and a host computer. 

During the experiment, the host computer executes the control algorithm developed in MATLAB, which generates digital control signals. These signals are converted to a 0-10V analog voltage by the PCIE-1810 and then pass through the amplifier module E-505.00, expanding the range to 0-100V.

The SGS detects the structural deformation caused by the deflection of the PFSM. The resistance change is converted to an analog voltage by the E-509.S3 servo module and subsequently to a 0-10V digital signal by the PCIE-1810. The host computer linearizes this signal to represent a deflection angle range of 2 mrad. As a result, closed-loop control is achieved.

Model identification has been completed in the work \cite{yang2024first}, and we adopt its identified parameters to establish the PFSM model.

To evaluate the tracking accuracy of the control algorithm, two performance metrics are chosen. One of metrics is the root mean square error (RMSE), defined as follows:
\begin{equation}
	RMSE=\sqrt{{\sum\limits_{k=1}^{K}{\left\| {{\mathbf{\theta }}_{d}}\left( k \right)-\mathbf{\theta }\left( k \right) \right\|_{2}^{2}}}/{\sum\limits_{k=1}^{K}{\left\| {{\mathbf{\theta }}_{d}}\left( k \right) \right\|_{2}^{2}}}}
	\label{eq:38}
\end{equation}

The other metric is the maximum absolute error (MAE), defined as:
\begin{equation}
	MAE=\underset{k}{\mathop{\max }}\,{{\left\| {{\mathbf{\theta }}_{d}}\left( k \right)-\mathbf{\theta }\left( k \right) \right\|}_{2}}
	\label{eq:39}
\end{equation}
where $K$ is the number of samples, ${{\mathbf{\theta }}_{d}}$ and $\mathbf{\theta }$ are the desired and actual output trajectories, respectively. ${{\left\| {{\mathbf{\theta }}_{d}}\left( k \right)-\mathbf{\theta }\left( k \right) \right\|}_{2}}$ represents the tracking error, ${{\left\| \cdot  \right\|}_{2}}$ denotes the ${{\ell }_{2}}$ norm. It should be noted that the execution time for input and output of the analog signals is approximately 0.3ms, whereas the control algorithm operates in 0.1ms. Consequently, the acquisition interval time is set as 0.5ms, indicating that the control bandwidth is 2kHz.

\subsection{State Comparison}
\subsubsection{Controllable Canonical Form vs. Non-Minimal State-Space}
There are two main methods for constructing discrete state-space equations from discrete transfer functions when applied to MPC. One approach is to convert the system into the controllable canonical form (CCF), where the states are typically not directly measurable.

Given the influence of process and observation noise, directly calculating the state vector $\mathbf{x}\left( k \right)$ from the previous state $\mathbf{x}\left( k-1 \right)$ in a forward manner or the observation $\mathbf{\theta }\left( k \right)$ in a backward manner are inaccurate. Therefore, a state observer is required for estimation.

Therefore, a Kalman filter \cite{kim2018introduction} is employed to compute the posterior state $\mathbf{\hat{x}}\left( k \right)$, which minimizes the error with respect to the true state and is used as the optimal estimation of $\mathbf{x}\left( k \right)$. It is assumed that the process noise and observation noise are independent and follow normal distributions, with covariance matrices denoted as $\bar{Q}$ and $\bar{R}$, respectively. For simplicity, it is assumed that $\bar{Q}$ and $\bar{R}$ are diagonal matrices with identical elements, whose choice of values influences the convergence of the Kalman filter.

An alternative approach utilizes the nonminimal state-space (NMSS) model \cite{wang2006improved}, in which the state variables are formed entirely from historical input and output data, eliminating the need for a state observer. However, its primary drawback is the increased dimensionality of the state variables. As a result, matrices may become excessively large for the same prediction horizon, leading to a substantial increase in computational complexity.

Using the discrete state-space equations derived from the two aforementioned methods, we compare the tracking accuracy based on the proposed controller. To more effectively assess the differences caused by various state construction methods, the error integral state $\mathbf{h}\left( k \right)$ is temporarily disabled, with the error integral gain matrix set as ${{K}_{I}}=0$. The reference trajectories for both the X-axis and Y-axis are selected as sine waves at 100 Hz, i.e., ${{\theta }_{d}}\left( t \right)=0.8\sin \left( 200\pi t \right)$, in units of $mrad$. For the CCF, the parameters that yield the minimum RMSE are $N=4$, $\rho ={{10}^{-8}}$, $\bar{Q}=200I$, and $\bar{R}=4\times {{10}^{3}}I$. For the NMSS, the parameters are $N=4$ and $\rho ={{10}^{-8}}$. The X-axis and Y-axis are tracking simultaneously, and this will be the case for all subsequent experiments.

\begin{figure}[htpb]
	\centering\includegraphics[width=0.9\columnwidth]{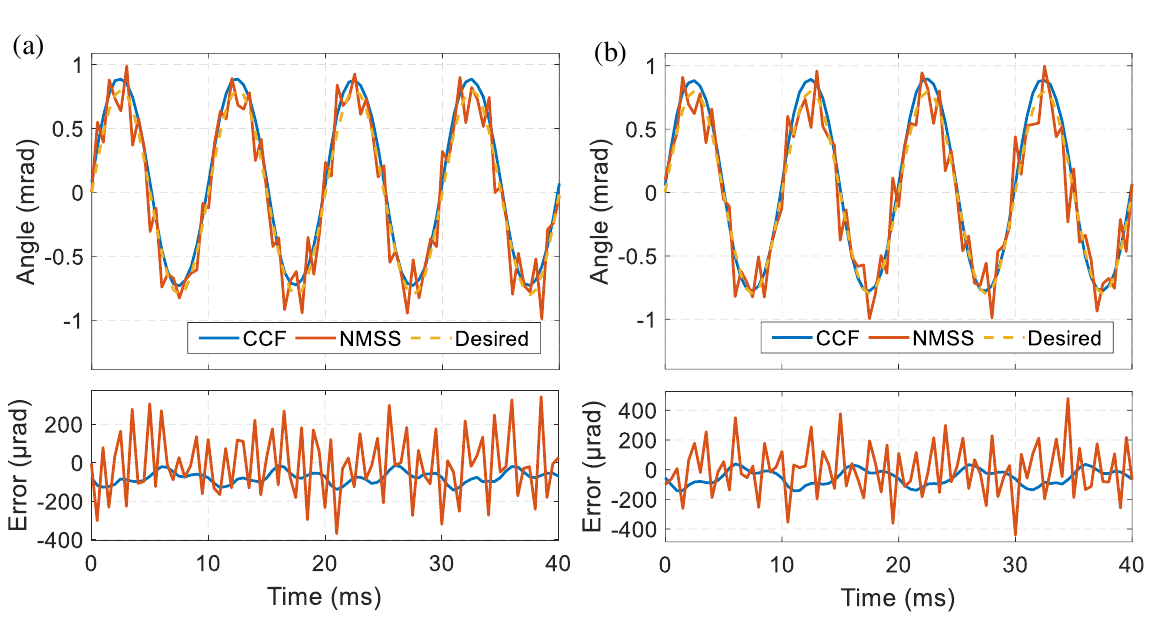}
	\caption{Trajectory tracking results and errors under different state-space equation construction methods. (a) X-axis. (b) Y-axis.}
	\label{fig:5}
\end{figure}

Fig. \ref{fig:5} presents the trajectory tracking results and errors of the dual axes under different discrete state-space equations. The RMSE and MAE of the tracking trajectories are provided in Tab. \ref{tab:1}.

It can be observed that the trajectory tracking accuracy of the X-axis and Y-axis using the CCF-based MPC is significantly better than that of the NMSS-based approach. The RMSE of the CCF method is reduced by 54.7\% compared to the NMSS method, and the MAE decreases by 347.71 $\mu rad$. Although the NMSS method does not require an additional state estimator, the process noise partly originates from the input, and the observation noise directly affects the output. This leads to a significant discrepancy between the true state and the state constructed by input-output. Since the MPC solution is a function of the state, the CCF-based method, which employs a Kalman filter to recover the true state from noise, achieves higher tracking accuracy.

\begin{table}[htpb!]
	\centering
	\caption{RMSE and MAE of trajectory tracking under different state-space equation construction methods}
	\label{tab:1}
	\begin{tabular}{@{}ccc@{}}
		\toprule
		State-space Construction & CCF  & NMSS \\ \midrule
		RMSE & \textbf{0.1396} & 0.3079  \\
		MAE$(\mu rad)$  & \textbf{196.88} & 544.59  \\ \bottomrule
	\end{tabular}
\end{table}

\subsubsection{MIMO Transfer Function with Creep vs. without Creep}
Compared to the NMSS, the system state constructed using the CCF reduces the dimensionality from 68 to 33, halving it and significantly decreasing the computational complexity of the MPC. Furthermore, there are 12 dimensions from the creep transfer function in the CCF, while 21 dimensions from the electromechanical transfer function. When tracking high-speed trajectories, the creep dynamic becomes inactive as it exhibits a constant response of one at high frequencies, permitting its neglect. Consequently, the linear transfer function matrix in Eq. (\ref{eq:6}) becomes equivalent to the electromechanical coupling matrix, thereby reducing the corresponding state dimensions to 21 and further decreasing the computational complexity. It should be noted that when tracking low-speed trajectories, more complex control strategies can be implemented to achieve ultra-high precision due to the sufficient computational time, which lies beyond the scope of this paper.

\begin{figure}[htpb]
	\centering\includegraphics[width=0.9\columnwidth]{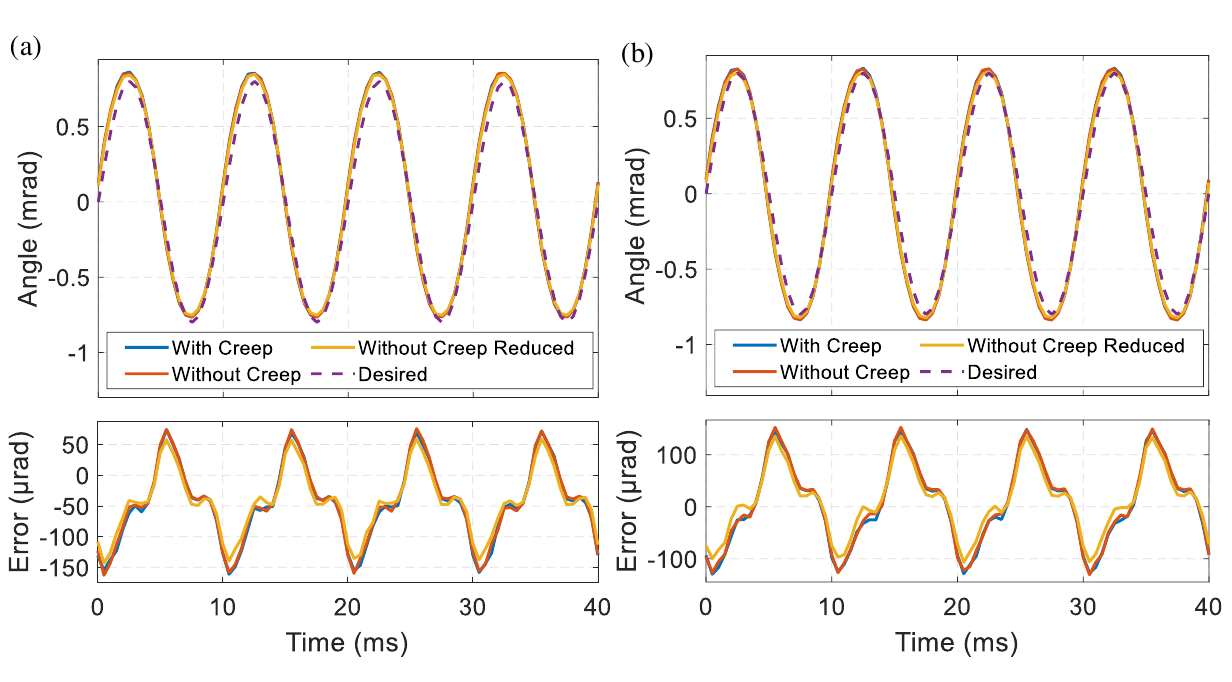}
	\caption{The MPC trajectory tracking results and errors for cases with and without creep in the transfer function matrix. (a) X-axis. (b) Y-axis.}
	\label{fig:6}
\end{figure}

In the case of the MIMO matrix with and without creep dynamics, we compared the tracking accuracy of the proposed method. The error integration state $\mathbf{h}\left( k \right)$ is similarly omitted by setting ${{K}_{I}}=0$. The high-speed reference trajectory remains a 100Hz sine wave, i.e., ${{\theta }_{d}}\left( t \right)=0.8\sin \left( 200\pi t \right)$, with units in mrad. For the Kalman filter design, the process noise covariance is set to $\bar{Q}=200I$, and the observation noise covariance is set to $\bar{R}=4\times {{10}^{3}}I$. The parameters $N=4$ and $\rho ={{10}^{-8}}$ are used for both system dynamics with or without creep. Fig. \ref{fig:6} presents the trajectory tracking results and errors of the dual axes under different transfer function matrices. The RMSE and MAE of the tracking trajectories are provided in Tab. \ref{tab:2}.

The trajectory tracking accuracy of the MPC without creep is slightly better than that with creep. This is because the poles and zeros of the discrete creep transfer function lie close to the unit circle. Although it is theoretically stable, truncation errors in numerical calculations may lead to divergence in state iteration estimates at some instants, thereby impacting tracking performance.

In addition to the improvement in accuracy, eliminating creep from the transfer function matrix reduces the system state dimensions from 33 to 21, leading to a computational load reduction of 59.5\% due to the complexity of $o({{n}^{2}})$. This represents a significant enhancement in efficiency and a reduction in costs. From another perspective, approximating the creep dynamics as one is equivalent to canceling all its poles and zeros. This raises the question of whether stable poles and zeros in the electromechanical dynamics can also be approximately canceled. Fortunately, we find that the single-axis transfer function contains two pairs of stable approximate poles and zeros that can be canceled, along with one pair in the cross-axis transfer function. Consequently, the system order is reduced, and the reidentified electromechanical transfer function matrices are as follows:
\begin{equation}
	\begin{aligned}
		& {{G}_{EM,XX}}=\frac{1.509\times {{10}^{13}}s+2.03\times {{10}^{12}}}{{{s}^{4}}+1.135\times {{10}^{8}}{{s}^{3}}+7.095\times {{10}^{11}}{{s}^{2}}+1.13\times {{10}^{15}}s+7.04\times {{10}^{8}}} \\ 
		& {{G}_{EM,YY}}=\frac{1.737\times {{10}^{14}}s+5.441\times {{10}^{13}}}{{{s}^{4}}+1.14\times {{10}^{9}}{{s}^{3}}+8.011\times {{10}^{12}}{{s}^{2}}+1.247\times {{10}^{16}}s+4.259\times {{10}^{11}}} \\ 
		& {{G}_{EM,XY}}=\frac{3.728\times {{10}^{4}}{{s}^{2}}-7.692\times {{10}^{6}}s+4.68\times {{10}^{11}}}{{{s}^{4}}+5.633\times {{10}^{4}}{{s}^{3}}+1.234\times {{10}^{9}}{{s}^{2}}+3.369\times {{10}^{12}}s+4.755\times {{10}^{15}}} \\ 
		& {{G}_{EM,YX}}=\frac{1.537\times {{10}^{5}}{{s}^{2}}+1.543\times {{10}^{8}}s+1.749\times {{10}^{12}}}{{{s}^{4}}+5.103\times {{10}^{4}}{{s}^{3}}+8.02\times {{10}^{8}}{{s}^{2}}+2.18\times {{10}^{12}}s+2.77\times {{10}^{15}}}
	\end{aligned}	
	\label{eq:40}
\end{equation}

Their frequency domain fitness is decreased by less than 2\% compared to the results in \cite{yang2024first}. With the same parameter settings of $N=4$ and $\rho ={{10}^{-8}}$, the tracking results based on the reduced-order electromechanical transfer function matrix are illustrated in Fig. \ref{fig:6}, while the RMSE and MAE are listed in Tab. \ref{tab:2}. It is evident that the tracking accuracy improves significantly after reduction, with the RMSE decreasing by 14.9\% and the MAE reducing by 27.9 $\mu rad$. In terms of computational complexity, the state dimension decreases from 21 to 16, representing a total reduction of up to 76\% compared to the transfer function matrix with creep. This indicates that the optimization is highly effective, enhancing tracking accuracy while substantially alleviating computational burdens, thus significantly increasing the upper limit of control bandwidth.
\begin{table}[htpb!]
	\centering
	\caption{RMSE and MAE of trajectory tracking under different transfer function matrices}
	\label{tab:2}
	\begin{tabular}{@{}cccc@{}}
		\toprule
		& With Creep    & Without Creep    & Without Creep Reduced \\ \midrule
		RMSE      & 0.1393 & 0.1386 & \textbf{0.1179} \\
		MAE$(\mu rad)$ & 203.84 & 202.48 & \textbf{174.54} \\ \bottomrule
	\end{tabular}
\end{table}

\subsubsection{ffectiveness of Error Integral State}
\begin{figure}[htpb]
	\centering\includegraphics[width=0.9\columnwidth]{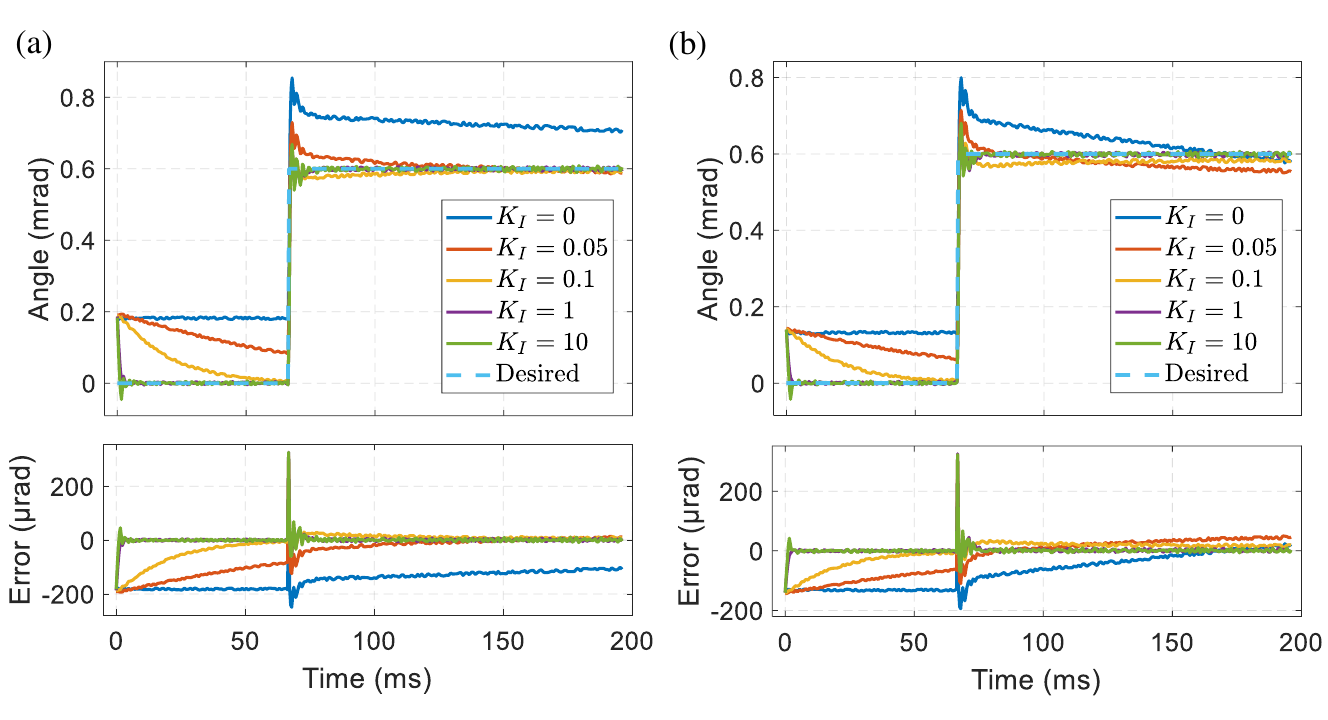}
	\caption{Trajectory tracking results and errors under different error integral coefficients ${{K}_{I}}$, where ${{K}_{I}}=0$ indicates the absence of error integral state. (a) X-axis. (b) Y-axis.}
	\label{fig:7}
\end{figure}

In this section, we evaluate the impact of the error integral state $\mathbf{h}\left( k \right)$ on tracking performance. A step signal is chosen as the reference trajectory, with the prediction horizon set at $N=4$. The trajectory tracking results and errors for different integral coefficients ${{K}_{I}}$ are presented in Fig. \ref{fig:7}, and the corresponding RMSEs are provided in Tab. \ref{tab:3}.

Since the open-loop zero position is manually tuned, an initial error is present. When ${{K}_{I}}=0$, the system state changes minimally, and the output voltage remains constant, resulting in a steady-state error that is equal to the initial error, with an overshoot of up to 316.4 $\mu rad$ following the step. As ${{K}_{I}}$ increases, the rate of error reduction prior to the step accelerates, and the overshoot following the step decreases rapidly, resulting in a significant reduction in RMSE. When ${{K}_{I}}=1$, the RMSE decreases by 94.2\% compared to that without error integral state (${{K}_{I}}=0$), while the overshoot decreases by 231.23 $\mu rad$. Therefore, the introduction of the error integral state effectively eliminates steady-state errors, thereby improving tracking performance. Moreover, for rapidly stabilizing coefficients ${{K}_{I}}=1$ and ${{K}_{I}}=10$, the ratio of ${{K}_{I}}$ to $\sqrt{\rho }$ equals ${{10}^{4}}$, which coincidentally corresponds to the magnitude of the ratio of control value to tracking error, thereby offering theoretical guidance for parameter tuning of the proposed MPC in other scenarios. Consequently, the highest tracking accuracy is achieved with $\left( {{K}_{I}}=1,\rho ={{10}^{-8}} \right)$, which is therefore designated as the optimal parameter set for subsequent comparisons in PFSM control.

\begin{table}[htpb!]
	\centering
	\caption{RMSE of trajectory tracking with different error integral coefficients ${{K}_{I}}$}
	\label{tab:3}
	\begin{tabular}{@{}cccccc@{}}
		\toprule
		$\left( {{K}_{I}},\rho  \right)$    & $\left( 0,{{10}^{-8}} \right)$      & $\left( 0.05,{{10}^{-8}} \right)$      & $\left( 0.1,{{10}^{-8}} \right)$     & $\left( 1,{{10}^{-8}} \right)$      & $\left( 10,{{10}^{-6}} \right)$      \\ \midrule
		RMSE & 0.2557 & 0.1548 & 0.093 & \textbf{0.0387} & 0.0419 \\ \bottomrule
	\end{tabular}
\end{table}

\subsection{Comparison with Baseline Methods}
In this section, we perform tracking control experiments using step signals, triangular signals, sinusoidal signals, and composite signals as desired trajectories. To validate the effectiveness of the proposed MPC combined with Inverse Bouc-Wen, we select the methods PID, DIM \cite{tang2010compensating}, FxVSNLMS \cite{wang2021laser}, and SPSASMC \cite{liu2023saturated}, which have demonstrated good performance, as baseline methods. It should be noted that all methods are evaluated without parallel PID controllers, allowing for a more direct comparison. Additionally, the methods PID, \cite{tang2010compensating}, and \cite{wang2021laser} require a saturation function similar to that embedded in \cite{liu2023saturated}; that is, if the input voltage exceeds the physical limits, it will be constrained to the upper or lower boundary values.

\subsubsection{Step Trajectory}
\begin{figure}[htpb]
	\centering\includegraphics[width=0.9\columnwidth]{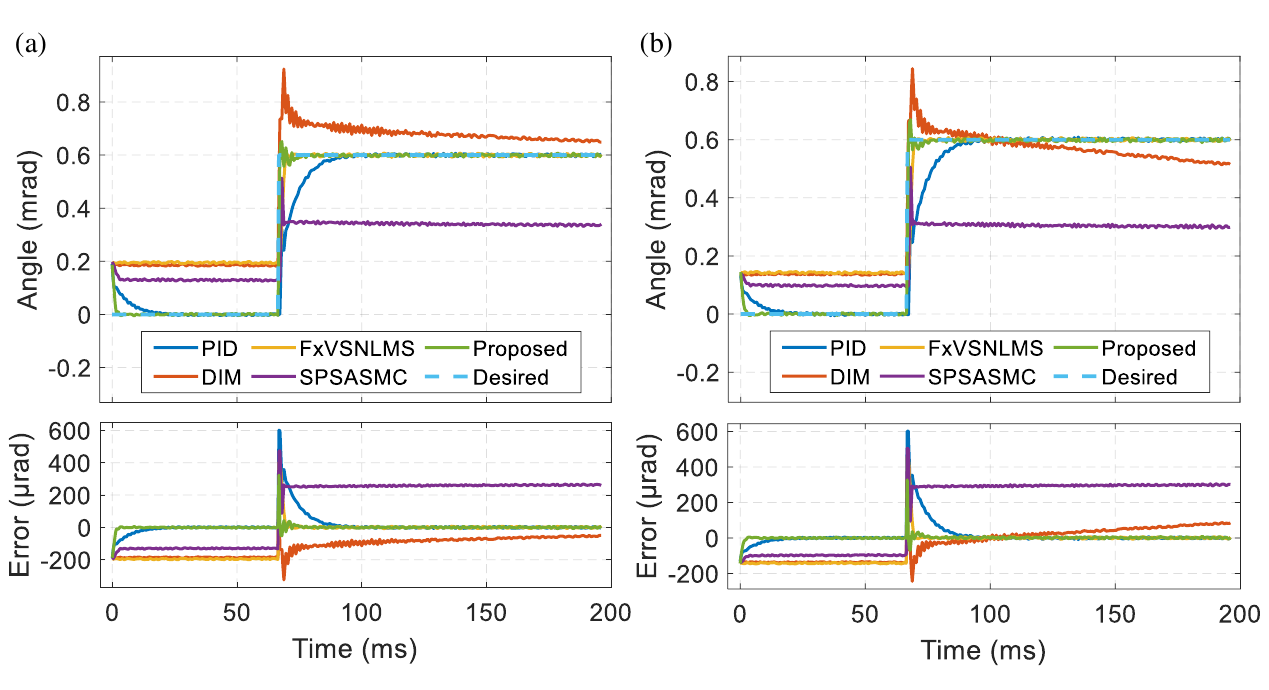}
	\caption{The step trajectory tracking results and errors of different control methods. (a) X-axis. (b) Y-axis.}
	\label{fig:8}
\end{figure}

Step responses are employed to compare the tracking performance of different control methods on discontinuous trajectories. The trajectory tracking results and errors for both axes are presented in Fig. \ref{fig:8}, with the corresponding RMSE and MAE detailed in Tab. \ref{tab:4}.

The PID controller exhibits no steady-state error due to the integral term; however, its rise time is 23 ms, with a control bandwidth of approximately 50 Hz, indicating insufficient tracking capability for high-speed trajectories. In the case of the open-loop DIM, the overshoot reaches 406.3 $\mu rad$, accompanied by transient high-frequency oscillations. The rise time for the FxVSNLMS is reduced to 4 ms; however, it exhibits a steady-state error when the initial desired trajectory is zero, as the tapped coefficients remain zero during this phase, thereby keeping the output voltage at its initial value. Although the SPSASMC exhibits a nonlinear relationship with the error, it does not include an error integral term, resulting in a steady-state error of 395.64 $\mu rad$ following the step change. The transient and steady-state performance of the aforementioned controllers are difficult to ensure simultaneously. In contrast, the proposed MPC achieves the shortest rise time of 1 ms, exhibits minimal overshoot, and maintains no steady-state error, resulting in a significant improvement in tracking accuracy. The RMSE is decreased by an average of 82.4\% compared to other methods, while the MAE is similarly reduced by an average of 22.8\%. Therefore, the proposed method exhibits optimal tracking performance on discontinuous trajectories.
\begin{table}[htpb!]
	\centering
	\caption{The RMSE and MAE of different control methods for tracking step trajectories}
	\label{tab:4}
	\begin{tabular}{@{}cccccc@{}}
		\toprule
		Control Methods      & PID    & \begin{tabular}[c]{@{}c@{}}DIM\\ \cite{tang2010compensating}\end{tabular}    & \begin{tabular}[c]{@{}c@{}}FxVSNLMS\\ \cite{wang2021laser}\end{tabular} & \begin{tabular}[c]{@{}c@{}}SPSASMC\\ \cite{liu2023saturated}\end{tabular} & Proposed MPC   \\ \midrule
		RMSE      & 0.1405 & 0.2336 & 0.2206   & 0.486   & \textbf{0.0391} \\
		MAE$(\mu rad)$ & 850.59 & 406.29 & 611.97   & 691.45  & \textbf{459.91} \\ \bottomrule
	\end{tabular}
\end{table}

\subsubsection{Sinusoidal Trajectory}
\begin{figure}[htpb]
	\centering\includegraphics[width=0.9\columnwidth]{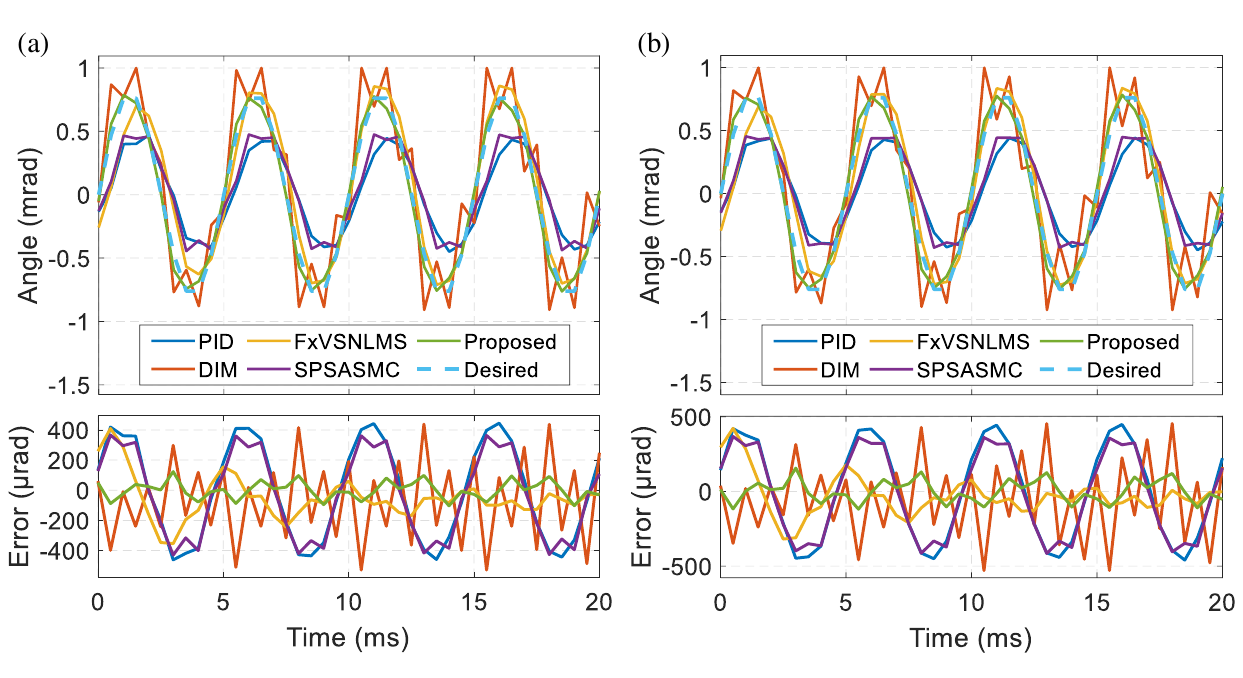}
	\caption{The 200 Hz sinusoidal trajectory tracking results and errors of different control methods. (a) X-axis. (b) Y-axis.}
	\label{fig:9}
\end{figure}

Sinusoidal trajectories are employed to evaluate the tracking performance of different control methods on smooth trajectories. To assess the robustness of different control methods to frequency variations, tracking experiments are conducted at 10 Hz, 50 Hz, 200 Hz, and 400 Hz. The RMSE and MAE of the tracking trajectories are presented in Tabs. \ref{tab:5} and \ref{tab:6}, respectively. The tracking results and tracking errors for the 200 Hz are illustrated in Fig. \ref{fig:9}. It should be noted that FxVSNLMS is a finite impulse response digital filter, and the tapped coefficients at the initial moment have not yet stabilized, leading to a larger tracking error. Therefore, tracking performance metrics are computed beginning from the second sine period.

For slow rate trajectories, the RMSE of controllers remains below 0.3 except for the SPSASMC, indicating good tracking accuracy. As the frequency increases, the tracking performance of PID and DIM deteriorates rapidly. When the frequency reaches 200 Hz, the RMSE for PID and DIM surpasses 0.5, presenting poor tracking accuracy. When the frequency reaches 400 Hz, the tap coefficient order of FxVSNLMS surpasses the number of sampling points in one sine period, causing significant phase delay and RMSE increase, leading to a sharp deterioration in tracking accuracy. In contrast, our proposed method maintains the RMSE below 0.2 at all frequencies, demonstrating exceptional tracking performance and robustness to frequency variations.

\begin{table}[htpb!]
	\centering
	\caption{RMSE of Control Methods for Tracking Sinusoidal Trajectories at Different Frequencies}
	\label{tab:5}
	\begin{tabular}{@{}cccccc@{}}
		\toprule
		Control Methods      & PID    & \begin{tabular}[c]{@{}c@{}}DIM\\ \cite{tang2010compensating}\end{tabular}    & \begin{tabular}[c]{@{}c@{}}FxVSNLMS\\ \cite{wang2021laser}\end{tabular} & \begin{tabular}[c]{@{}c@{}}SPSASMC\\ \cite{liu2023saturated}\end{tabular} & Proposed MPC               \\ \midrule
		10Hz  & 0.2956 & 0.1991 & 0.0792 & 0.4319 & \textbf{0.0109} \\
		50Hz  & 0.4759 & 0.2982 & 0.1069 & 0.4655 & \textbf{0.0313} \\
		200Hz & 0.5717 & 0.5086 & 0.2652 & 0.5093 & \textbf{0.1174} \\
		400Hz & 0.6451 & 0.5499 & 0.5446 & 0.5559 & \textbf{0.1977} \\ \bottomrule
	\end{tabular}
\end{table}

\begin{table}[htpb!]
	\centering
	\caption{MAE ($\mu rad$) of Control Methods for Tracking Sinusoidal Trajectories at Different Frequencies}
	\label{tab:6}
	\begin{tabular}{@{}cccccc@{}}
		\toprule
		Control Methods      & PID    & \begin{tabular}[c]{@{}c@{}}DIM\\ \cite{tang2010compensating}\end{tabular}    & \begin{tabular}[c]{@{}c@{}}FxVSNLMS\\ \cite{wang2021laser}\end{tabular} & \begin{tabular}[c]{@{}c@{}}SPSASMC\\ \cite{liu2023saturated}\end{tabular} & Proposed MPC              \\ \midrule
		10Hz  & 347.05 & 314.47 & 127.37 & 521.02 & \textbf{17.84}  \\
		50Hz  & 549.32 & 523.96 & 140.71 & 557.19 & \textbf{52.91}  \\
		200Hz & 642.98 & 748.51 & 588.07 & 589.05 & \textbf{199.31} \\
		400Hz & 819.17 & 1068.4 & 960.61 & 746.74 & \textbf{319.47} \\ \bottomrule
	\end{tabular}
\end{table}

\begin{figure}[htpb!]
	\centering\includegraphics[width=0.9\columnwidth]{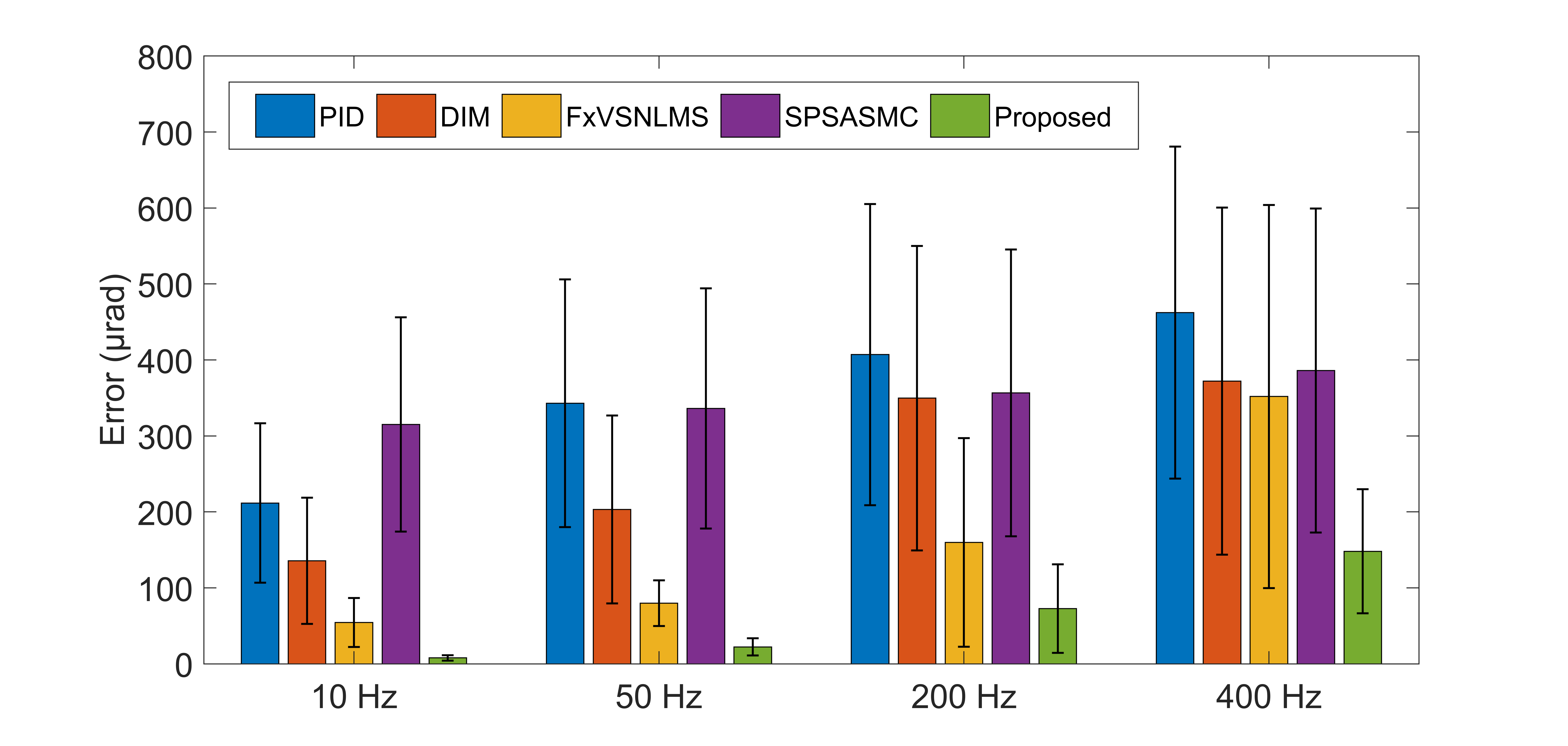}
	\caption{Statistical histograms of sinusoidal trajectories tracking errors at various frequencies for different control methods. The solid bars represent the mean tracking errors, while the error bars indicate the variance.}
	\label{fig:10}
\end{figure}

Furthermore, statistical histograms of the tracking errors for each control method are performed as shown in Fig. \ref{fig:10}. The SPSASMC exhibits relatively large means and variances in tracking errors across all frequencies. As the frequency increases, the means and variances of tracking errors for PID, DIM, and FxVSNLMS increase substantially, reflecting deteriorating tracking performance. In contrast, the proposed method consistently maintains low mean and variance in tracking errors across all frequencies, highlighting its exceptional robustness to frequency variations.

\subsubsection{Triangular Trajectory}
\begin{figure}[htpb]
	\centering\includegraphics[width=0.9\columnwidth]{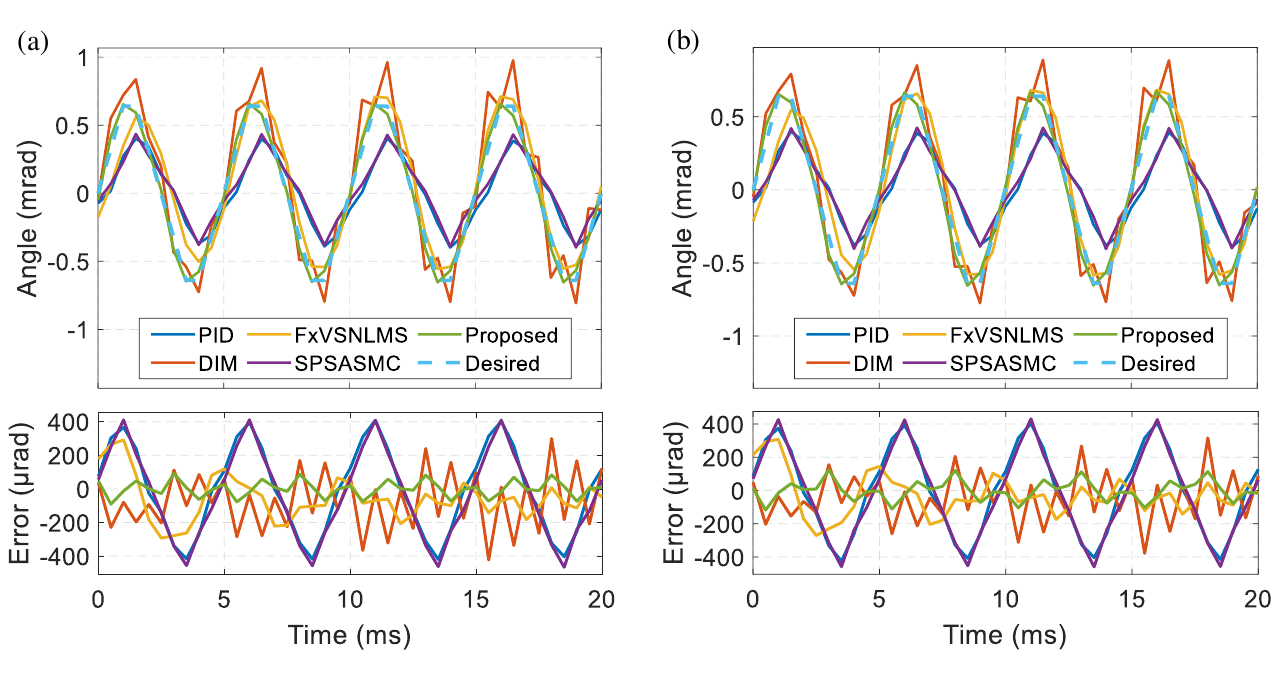}
	\caption{The 200 Hz triangular trajectory tracking results and errors of different control methods. (a) X-axis. (b) Y-axis.}
	\label{fig:11}
\end{figure}

Triangular trajectories are employed to evaluate the tracking performance of different control methods on non-smooth trajectories at 10 Hz, 50 Hz, 200 Hz, and 400 Hz. The RMSE and MAE of the tracking trajectories are presented in Tabs. \ref{tab:7} and \ref{tab:8}, respectively. The tracking results and tracking errors for the 200 Hz are depicted in Fig. \ref{fig:11}.

The methods, except SPSASMC, achieve satisfactory tracking accuracy for low-rate trajectories. As the frequency increases, the tracking accuracy of PID, DIM, and FxVSNLMS declines rapidly, resulting in trajectory distortion. In contrast, our proposed method consistently tracks the desired trajectory across a wide range of frequencies.

\begin{table}[htpb!]
	\centering
	\caption{RMSE of Control Methods for Tracking Triangular Trajectories at Different Frequencies}
	\label{tab:7}
	\begin{tabular}{@{}cccccc@{}}
		\toprule
		Control Methods      & PID    & \begin{tabular}[c]{@{}c@{}}DIM\\ \cite{tang2010compensating}\end{tabular}    & \begin{tabular}[c]{@{}c@{}}FxVSNLMS\\ \cite{wang2021laser}\end{tabular} & \begin{tabular}[c]{@{}c@{}}SPSASMC\\ \cite{liu2023saturated}\end{tabular} & Proposed MPC              \\ \midrule
		10Hz  & 0.307  & 0.2423 & 0.0959 & 0.4952 & \textbf{0.0144} \\
		50Hz  & 0.4796 & 0.3091 & 0.1662 & 0.5365 & \textbf{0.0385} \\
		200Hz & 0.5793 & 0.3969 & 0.3051 & 0.5862 & \textbf{0.1207} \\
		400Hz & 0.7268 & 0.4522 & 0.5838 & 0.7009 & \textbf{0.2275} \\ \bottomrule
	\end{tabular}
\end{table}

\begin{table}[htpb!]
	\centering
	\caption{MAE $(\mu rad)$ of Control Methods for Tracking Triangular Trajectories at Different Frequencies}
	\label{tab:8}
	\begin{tabular}{@{}cccccc@{}}
		\toprule
		Control Methods      & PID    & \begin{tabular}[c]{@{}c@{}}DIM\\ \cite{tang2010compensating}\end{tabular}    & \begin{tabular}[c]{@{}c@{}}FxVSNLMS\\ \cite{wang2021laser}\end{tabular} & \begin{tabular}[c]{@{}c@{}}SPSASMC\\ \cite{liu2023saturated}\end{tabular} & Proposed MPC             \\ \midrule
		10Hz  & 254.42 & 318.49 & 113.54 & 561.77 & \textbf{24.28}  \\
		50Hz  & 490.03 & 304.89 & 244.14 & 625.53 & \textbf{59.76}  \\
		200Hz & 594.38 & 567.24 & 423.95 & 652.63 & \textbf{154.31} \\
		400Hz & 873.68 & 610.71 & 760.96 & 767.8  & \textbf{255.11} \\ \bottomrule
	\end{tabular}
\end{table}

\begin{figure}[htpb!]
	\centering\includegraphics[width=0.9\columnwidth]{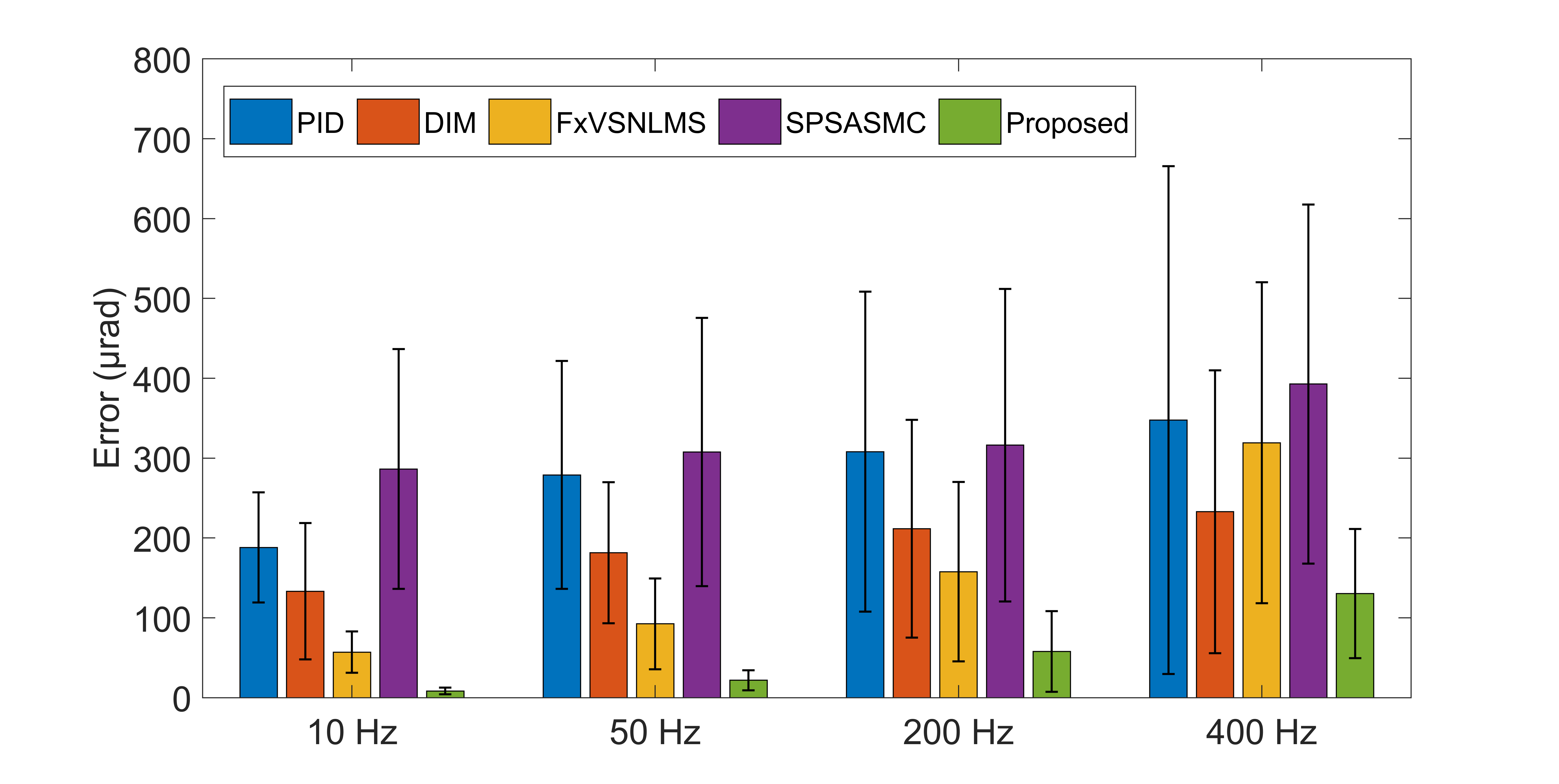}
	\caption{Statistical histograms of triangular trajectories tracking errors at various frequencies for different control methods. The solid bars represent the mean tracking errors, while the error bars indicate the variance.}
	\label{fig:12}
\end{figure}

Similarly, statistical histograms of the tracking errors are presented in Fig. \ref{fig:12}. As the frequency increases, the means and variances of tracking errors for PID, DIM, and FxVSNLMS increase significantly. In contrast, the proposed method consistently maintains low means and variances across various frequencies.

\subsubsection{Composite trajectory}
\begin{figure}[htpb!]
	\centering\includegraphics[width=0.9\columnwidth]{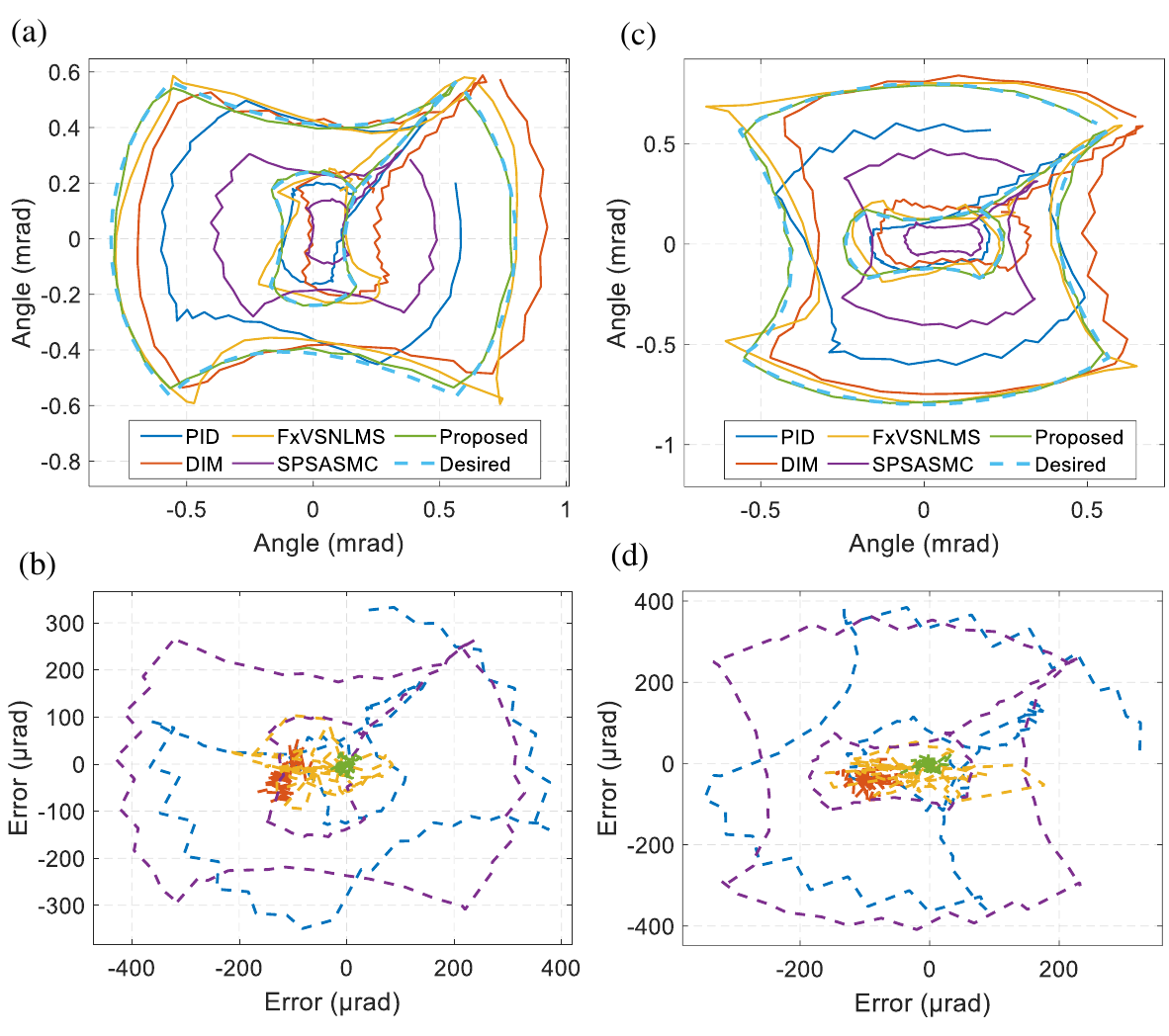}
	\caption{The composite trajectory tracking results and errors of different control methods. (a) $({{\theta }_{d,1}},{{\theta }_{d,2}})$ as reference trajectories for the X-axis and Y-axis, respectively, and (b) corresponding tracking errors. (c) $({{\theta }_{d,2}},{{\theta }_{d,1}})$ as reference trajectories for the X-axis and Y-axis, respectively, and (d) corresponding tracking errors.}
	\label{fig:13}
\end{figure}

Furthermore, composite trajectory tracking experiments are conducted. ${{\theta }_{d,1}}$ and ${{\theta }_{d,2}}$ consist of a ramp and a 100 Hz sinusoidal signal. The tracking results and errors for the X-axis and Y-axis are presented in Fig. \ref{fig:13}. The RMSE and MAE of both tracking trajectories are provided in Tab. \ref{tab:9}.

\begin{table}[h!]
	\centering
	\caption{RMSE and MAE ($\mu rad$) of Different Control Methods for Tracking Composite Trajectories}
	\label{tab:9}
	\begin{tabular}{@{}ccccccc@{}}
		\toprule
		\multicolumn{2}{c}{Control Methods}                                                      & PID    & \begin{tabular}[c]{@{}c@{}}DIM\\ \cite{tang2010compensating}\end{tabular}    & \begin{tabular}[c]{@{}c@{}}FxVSNLMS\\ \cite{wang2021laser}\end{tabular} & \begin{tabular}[c]{@{}c@{}}SPSASMC\\ \cite{liu2023saturated}\end{tabular} & Proposed MPC              \\ \midrule
		\multirow{2}{*}{RMSE}                                                 & $\left( {{\theta }_{d,1}},{{\theta }_{d,2}} \right)$ & 0.4227 & 0.2277 & 0.1673 & 0.4885 & \textbf{0.0298} \\
		& $\left( {{\theta }_{d,2}},{{\theta }_{d,1}} \right)$ & 0.4224 & 0.2046 & 0.1458 & 0.4893 & \textbf{0.0293} \\ \midrule
		\multirow{2}{*}{\begin{tabular}[c]{@{}c@{}}MAE\\ $(\mu rad)$\end{tabular}} & $\left( {{\theta }_{d,1}},{{\theta }_{d,2}} \right)$ & 402.03 & 177.03 & 223.08 & 436.28 & \textbf{34.15}  \\
		& $\left( {{\theta }_{d,2}},{{\theta }_{d,1}} \right)$ & 402.77 & 152.6  & 184.89 & 436.82 & \textbf{50.95}  \\ \bottomrule
	\end{tabular}
\end{table}

The SPSASMC causes the 2D tracking trajectory to shrink in scale due to the fixed gain, aligning the error trajectory closely with the desired trajectory in shape. The PID also exhibits phase lags between the X-axis and Y-axis, causing a rotational transformation of error trajectory relative to the SPSASMC. Additionally, the error trajectory of the DIM is more concentrated with improved tracking accuracy attributable to its high-precision modeling. However, due to the open-loop control, the center of its error trajectory does not lie at (0,0). Although the error of FxVSNLMS is more dispersed compared to DIM, its center lies closer to (0,0) under closed-loop control, resulting in better tracking accuracy than DIM. Notably, the error trajectory of our proposed method is even more concentrated than DIM, as well as its center remains near (0,0) under closed-loop control. The RMSE of proposed method is significantly reduced by 93\%, 87\%, 82\%, and 94\% compared to PID, DIM, SPSASMC, and FxVSNLMS, respectively. The MAE is significantly reduced by 92\%, 81\%, 85\%, and 92\% compared to PID, DIM, SPSASMC, and FxVSNLMS, respectively, achieving superior tracking performance among control methods.

\section{Conclusion}
This paper proposes an MPC method that incorporates an inverse hysteresis to track the reference trajectories for the PFSM. Leveraging the convexity of the dual problem in quadratic programming, a coordinate descent algorithm is employed to efficiently solve the constrained optimization problem.

A PFSM trajectory tracking experimental platform is set up. Comparing the proposed MPC method with traditional model-free PID, and existing model-based open-loop DIM, closed-loop FxVSNLMS and SPSASMC controllers. For step response, the MPC method stabilizes within 0.8 ms with zero steady-state error, achieving an average RMSE reduction of 82.4\% compared to other methods. For continuous sinusoidal and triangular trajectories, the MPC demonstrates excellent tracking performance at various frequencies. At the low frequency of 10 Hz, the RMSE is significantly reduced by an average of 93.7\% compared to other methods, while reduced by an average of 68.9\% at the high frequency of 400 Hz. Further composite trajectory tracking experiments also validate this demonstration. The results provide valuable references for engineering applications.

A PFSM trajectory tracking experimental platform has been established. The proposed MPC method is compared with the traditional model-free PID, and existing model-based DIM with open-loop, as well as FxVSNLMS and SPSASMC with closed-loop. For step response, the MPC method stabilizes within 0.8 ms with zero steady-state error, achieving an average RMSE reduction of 82.4\% compared to other methods. For continuous sinusoidal and triangular trajectories, the MPC achieves superior tracking performance across a range of frequencies. At the low frequency of 10 Hz, the RMSE is significantly reduced by an average of 93.7\% compared to other methods, and by an average of 68.9\% at the high frequency of 400 Hz. Further composite trajectory tracking experiments also validate these findings. The results offer valuable insights for engineering applications.

\section*{Disclosures}
The authors declare no conflicts of interest.

\bibliographystyle{unsrtnat}
\bibliography{references}  %%% Uncomment this line and comment out the ``thebibliography'' section below to use the external .bib file (using bibtex) .




% 	\bibitem{keshet2016prediction}
% 	Keshet, Renato, Alina Maor, and George Kour.
% 	\newblock Prediction-Based, Prioritized Market-Share Insight Extraction.
% 	\newblock In {\em Advanced Data Mining and Applications (ADMA), 2016 12th International 
%                       Conference of}, pages 81--94,2016.

% \end{thebibliography}

\end{document}